\documentclass[iop,numberedappendix,apj]{emulateapj}
\pdfoutput=1
\usepackage{color}
\usepackage{amssymb}
\usepackage{natbib}
\usepackage{graphicx}
\usepackage{url}
\usepackage[tbtags]{amsmath}
\usepackage{hyperref,xcolor}
\hypersetup{colorlinks,linkcolor={blue!50!black},citecolor={blue!50!black},urlcolor={blue!80!black}}
\newcommand {\apgt} {\ {\raise-.5ex\hbox{$\buildrel>\over\sim$}}\ }
\newcommand {\aplt} {\ {\raise-.5ex\hbox{$\buildrel<\over\sim$}}\ }

\newcommand{\asec}{\mbox{$^{\prime\prime}$}}
\newcommand{\amin}{$^{\prime}$}

\newcommand{\etal}{{\it et al.}}

\newcommand{\xmm}{{XMM-{\it Newton}}}

\slugcomment{}
\shortauthors{Romero \etal}
\shorttitle{Joint SZE Map Fitting with MUSTANG and Bolocam}

\begin{document}

\title{Galaxy Cluster Pressure Profiles as Determined by Sunyaev Zel'dovich Effect 
  Observations with MUSTANG and Bolocam I: Joint Analysis Technique}
\author{
  Charles E. Romero\altaffilmark{1,2,3},
  Brian S. Mason\altaffilmark{2},
  Jack Sayers\altaffilmark{4}
  Alexander H.\ Young\altaffilmark{5},
  Tony Mroczkowski\altaffilmark{6,7},
  Tracy E. Clarke\altaffilmark{7},
  Craig Sarazin\altaffilmark{2},
  Jonathon Sievers\altaffilmark{8},
  Simon R. Dicker\altaffilmark{5},
  Erik D.\ Reese\altaffilmark{9},
  Nicole Czakon \altaffilmark{4,10},
  Mark Devlin\altaffilmark{5},
  Phillip M.\ Korngut\altaffilmark{4},
  Sunil Golwala\altaffilmark{4}
} 
\date{\today}

\altaffiltext{1}{National Radio Astronomy Observatory, 520 Edgemont
  Rd., Charlottesville VA 22903, USA} 
\altaffiltext{2}{Department of Astronomy, University of Virginia,
  P.O. Box 400325, Charlottesville, VA 22901, USA}
\altaffiltext{3}{Author contact: \email{cer2te@virginia.edu}}
\altaffiltext{4}{Department of Physics, Math, and Astronomy,
  California Institute of Technology, Pasadena, CA 91125, USA}
\altaffiltext{5}{Department of Physics and Astronomy, University of
  Pennsylvania, 209 South 33rd Street, Philadelphia, PA, 19104, USA}
\altaffiltext{6}{National Research Council Fellow} 
\altaffiltext{7}{U.S.\ Naval Research Laboratory,
  4555 Overlook Ave SW, Washington, DC 20375, USA}
\altaffiltext{8}{Astrophysics \& Cosmology Research Unit, University of KwaZulu-Natal,
  Private Bag X54001, Durban 4000, South Africa}
\altaffiltext{9}{Department of Physics, Astronomy, and Engineering, 
  Moorpark College, 7075 Campus Rd., Moorpark, CA 93021, USA} 
\altaffiltext{10}{Academia Sinica, 128 Academia Road, Nankang, Taipei 115, Taiwan}


\begin{abstract}
We present a technique to constrain galaxy cluster pressure profiles by jointly fitting  Sunyaev-Zel'dovich 
effect (SZE) data obtained with MUSTANG and Bolocam for the clusters Abell 1835 and MACS0647. Bolocam and 
MUSTANG probe different angular scales and are thus highly complementary. We find that the addition of the
high resolution MUSTANG data can improve constraints on pressure profile parameters relative to those 
derived solely from Bolocam. In Abell 1835 and MACS0647, we find gNFW inner slopes of 
$\gamma = 0.36_{-0.21}^{+0.33}$ and $\gamma = 0.38_{-0.25}^{+0.20}$, respectively when $\alpha$ and $\beta$ are
constrained to $0.86$ and $4.67$ respectively. The fitted SZE pressure profiles are in good agreement 
with X-ray derived pressure profiles. 
\end{abstract}


\maketitle

\section{Introduction}
\label{sec:intro}

Galaxy clusters are the largest gravitationally bound objects in the universe and thus serve as ideal cosmological probes 
and astrophysical laboratories. Because the formation of galaxy clusters stems from overdensities of matter
and depends on the cosmic composition of the universe, one can constrain cosmological parameters such as the matter 
density of the universe, $\Omega_m$, the matter power spectrum normalization, $\sigma_8$,
and the equation of state for dark energy density $\Omega_{\Lambda}$, $w$ \citep[e.g.][]{carlstrom2002}.

Within a galaxy cluster, the gas in the intracluster medium (ICM) constitutes 90\% of the
baryonic mass \citep{vikhlinin2006b} and is directly observable in the X-ray due to bremsstrahlung emission. 
At millimeter and sub-millimeter wavelengths, the ICM is observable via the Sunyaev-Zel'dovich effect (SZE) 
\citep{sunyaev1972}: the inverse Compton scattering of cosmic microwave background (CMB) photons off of
the hot ICM electrons. The thermal SZE is observed as an intensity decrement relative to the CMB at wavelengths longer 
than $\sim$1.4 mm (frequencies less than $\sim$220 GHz).
At longer radio wavelengths, if relativistic electrons are present, parts of the ICM may emit synchrotron emission.

In the core of a galaxy cluster, baryonic physics are non-negligible and non-trivial. Some observed
physical processes in the core include shocks and cold fronts \citep[e.g.][]{markevitch2007}, sloshing
\citep[e.g.][]{fabian2006}, and X-ray cavities \citep{mcnamara2007}. It is also theorized that helium sedimentation
should occur, most noticeably in low redshift, dynamically-relaxed clusters \citep{abramopoulos1981, gilfanov1984} 
and recently the expected helium enhancement via sedimentation has been numerically simulated \citep{peng2009}. 
This would result in an offset between X-ray and SZE derived pressure profiles.

At large radii ($R \gtrsim R_{500}$),\footnote{$R_{500}$
is the radius at which the enclosed average mass density is 500 times the critical density, 
$\rho_c(z)$, of the universe} equilibration timescales are longer, accretion is ongoing, 
and hydrostatic equilibrium (HSE) is a poor approximation. 
Several numerical simulations show that the fractional contribution
 from non-thermal pressure increases with radius \citep{shaw2010,battaglia2012,nelson2014}. 
For all three studies, non thermal pressure fractions between 15\% and 30\% are found at ($R \sim R_{500}$)
for redshifts $0 < z < 1$. However, the intermediate radii, between the core and outer regions of the 
galaxy cluster, offer a region where self-similar scalings derived from HSE can be used to describe simulations 
and observations \citep[e.g.][]{kravtsov2012}. Moreover, both simulations and observations find low
cluster-to-cluster scatter in pressure profiles within this intermediate radial range \citep[e.g.][]{borgani2004,
nagai2007,arnaud2010,bonamente2012,planck2013a,sayers2013}.

While many telescopes capable of making SZE observations are already operational or are being built, most have
angular resolutions (full width, half maximum - FWHM) of one arcminute or larger. The MUSTANG instrument \citep{dicker2008}
on the 100 meter Robert C. Byrd Green Bank Telescope \citep[GBT, ][]{jewell2004} with its angular resolution of $9\asec$ 
(FWHM) and sensitivity up to the limit of MUSTANG's instantaneous field of view, $1$\amin, 
is one of only a few SZE instruments with sub-arcminute resolution.
To probe a wider range of scales we complement our MUSTANG data with SZE data from Bolocam \citep{glenn1998}. 
Bolocam is a 144-element bolometer
array on the Caltech Submillimeter Observatory (CSO) with a beam FWHM of $58\asec$ at 140 GHz and circular FOV with $8'$ 
diameter, which is well matched to the angular size of $R_{500}$ ($\sim 4$\amin) for both of the clusters in our sample. 

In this paper, we extend the map fitting technique used in \citet{young2014}, to simultaneously
fit 3D pressure profiles to Bolocam and MUSTANG data. With MUSTANG's
high resolution, this is the first analysis to make use of SZE observations that cover similar scales 
($0.03 R_{500} <r \lesssim R_{500}$) to those probed by X-ray studies
\citep[][; hereafter N07 and A10 respectively]{nagai2007,arnaud2010} which have constrained the average cluster pressure profile. 
N07 compared X-ray and simulation results over radial scales ($0.1 R_{500} \lesssim r \lesssim R_{500} $), whereas
A10 used X-ray determined pressure profiles for $0.03 R_{500} \lesssim r < R_{500}$, and simulation results for
$R_{500} < r$. More recently, the Planck collaboration has published an analysis combining \xmm\ observations,
which span ranges $0.02 R_{500} < r < R_{500}$ with \emph{Planck} observations, which span radial ranges 
$0.1 R_{500} < r < 3 R_{500}$ \citep{planck2013a}. Additionally, a sample of clusters studied by Bolocam  
has been analyzed using solely the SZE data, which spans radial ranges of $0.07 R_{500} < r < 3.5 R_{500}$
\citep{sayers2013}.

This paper is organized as follows. In Section \ref{sec:obs} we describe the MUSTANG and Bolocam observations and reduction. 
In Section \ref{sec:jointfitting} we detail the method used to jointly fit pressure profiles to MUSTANG and Bolocam data. We
present results from the joint fits in Section \ref{sec:results} and discuss our results in Section \ref{sec:discussion}. 
Throughout this paper we assume a flat $\Lambda$CDM cosmology with $\Omega_m = 0.3$, $\Omega_{\lambda} = 0.7$, and $H_0 = 70$ 
km s$^{-1}$ Mpc$^{-1}$, consistent with the 9-year \emph{Wilkinson Microwave Anisotropy Probe} (WMAP) results reported in
 \cite{hinshaw2013}.


\section{Observations and Data Reduction}
\label{sec:obs}


\subsection{Sample}

To test the application of the joint fitting technique, we use two clusters: 
MACS J0647.4+7015 and Abell 1835. Properties of these two clusters are shown in 
Table \ref{table:clusterprops}. MACS 0647 \citep{ebeling2007,ebeling2010} has been observed 
by MUSTANG as part of an SZE program to observe the Cluster Lensing and Supernova with Hubble (CLASH) 
sample \citep{postman2012}. MACS J0647 is well detected by MUSTANG and appears to be relaxed;
thus it presents a good baseline for testing our fitting procedure. We note that it has been previously
analyzed \citep{young2014}, albeit with a slightly different approach.
Abell 1835 is a well known cool core cluster \citep[e.g.][]{peterson2001}, with prior SZE detections
\citep[e.g.][]{reese2002,benson2004,bonamente2006,sayers2011,mauskopf2012}. MUSTANG detects the cluster
along with a point source in the center of the cluster, thus making it a good cluster to demonstrate that the 
joint fitting technique established here can distinguish between point source signal and SZE signal. 

\begin{deluxetable}{ccccc}
\tabletypesize{\footnotesize}
\tablecolumns{5}
\tablewidth{0pt} 
\tablecaption{Cluster Properties \label{table:clusterprops}}
\tablehead{
\colhead{Cluster} & \colhead{$z$} & \colhead{$R_{500}$} & \colhead{$R_{500}$} & \colhead{$M_{500}$} \\
\colhead{}        & \colhead{}    & \colhead{(Mpc)}     & \colhead{(\arcmin)} & \colhead{($10^{14}~M_{\odot}$)} 
}
\startdata
    A1835        & 0.253 & $1.49 \pm 0.06$ & $6.30$ & $12.3 \pm 1.4$ \\
    MACS J0647.7 & 0.591 & $1.26 \pm 0.06$ & $3.16$ & $10.9 \pm 1.6$ 
\enddata
\tablerefs{\citet{mantz2010}}
\end{deluxetable}

\subsection{MUSTANG Observations}
\label{sec:musobs}

MUSTANG is a 64 pixel array of Transition Edge Sensor (TES) bolometers arranged in an $8 \times 8$ array
located at the Gregorian focus on the 100 m GBT. Operating at 90 GHz (81--99~GHz),
MUSTANG has an angular resolution of $9\asec$ and pixel spacing of 0.63$f \lambda$ resulting in a FOV
of $42\asec$. More detailed information about the instrument can be found in \citet{dicker2008}.

\begin{figure}[h]
\begin{center}
 \includegraphics[width=0.48\textwidth]{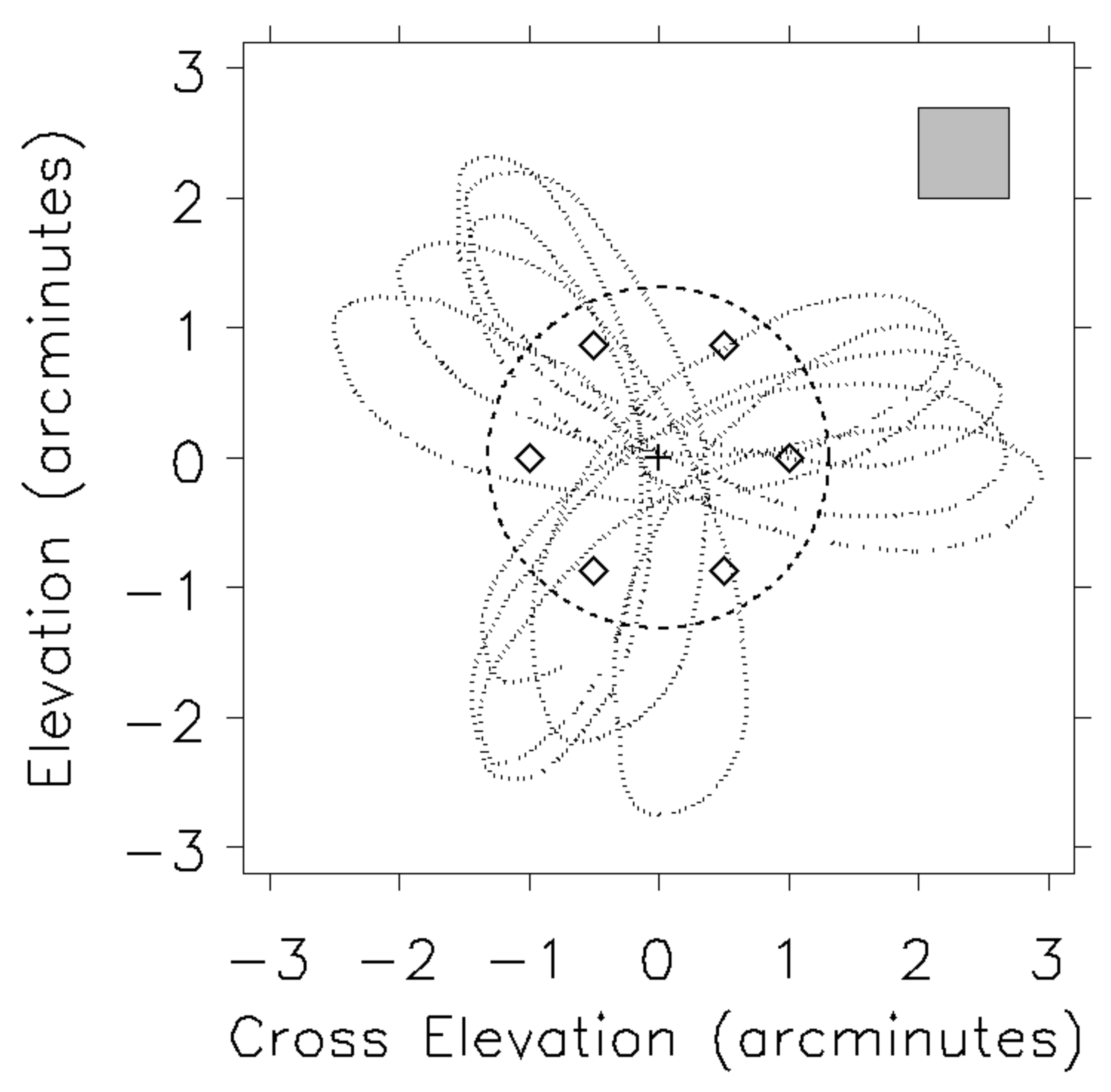}
\end{center}  
\caption{Dotted line: an example GBT trajectory for a 140 second scan with a Lissajous daisy scan pattern. 
    Three 140 second scans result in complete coverage of a circle with $3'$ radius. The plus sign 
    indicates the pointing center used for this daisy scan. The diamonds indicate six offset 
    pointings, used on observations of MACS 0647. The FOV of MUSTANG is shown by the shaded 
    box. The dashed circle encloses 50\% of the peak weight over the ensemble of scans for a given cluster.}
  \label{fig:daisy_traj}
\end{figure}

Observations consisted of scanning on pointing centers covering the central region of each galaxy cluster. 
A variety of scan durations ranging between 90 and 300 seconds were employed. Over the seasons of observations,
we found that scans lasting 200 to 220 seconds provided the best yields, and accordingly we began to favor 
scans of this length. Abell 1835 was observed in the winter/spring of 2009 and 2010 with one 
central pointing as described in \citet{korngut2011}. For MACS 0647 (observed 2011-2013), 
we adopted an observing strategy that consisted of one central pointing,
and six off-center pointings, spaced hexagonally such that each was $1\arcmin$ from the center, and $1\arcmin$ away from
the other two pointings (Figure \ref{fig:daisy_traj}). Using the offset pointings with a Lissajous daisy scan 
pattern provides more uniform coverage within the central arcminute of our maps than using only one central 
pointing with Lissajous daisy scans. For each pointing center, the Lissajous daisy has a $3\arcmin$ radius. The 
coverage (weight) drops to 50\% of its peak value at a radius of $1.3'$.

Absolute flux calibrations are based on the planets Mars, Uranus, or Saturn, nebulae, or the star Betelgeuse 
($\alpha_{Ori}$). At least one of these flux calibrators was observed at least once per night. 
Planets and nebulae are the preferred targets, as they can be
calibrated directly to WMAP observations \citep{weiland2011}. Betelgeuse can then be cross calibrated to
these planets and nebulae, and may be used as an absolute calibration itself when no planets or
nebulae were observed in a given night. We find our calibration is accurate to a
10\% RMS uncertainty.

At the start of each night of observing, medium-scale, mostly thermal imperfections in the
GBT surface are measured and corrected using out-of-focus (OOF) holographic technique
\citep{nikolic2007}.
Interspersed with scans on clusters, we observe nearby compact quasars as secondary calibrators. 
Secondary calibrators are observed roughly once every 30 minutes,
and allow us to track the pointing, beam profile, and gain
changes of the telescope. If the the beam shape or gain degrade by more than 10\%, another 
OOF measurement was performed. We find that our pointing accuracy is $2\asec$.
We pair the scans on secondary calibrators with off-source scans of
an internal calibration lamp (CAL) chopped with a 0.5 Hz square wave pattern. These CAL scans are taken
with the telescope at rest.

All observations of calibrators (primary and secondary) also make use of a Lissajous daisy scan
pattern. These scans have a duration of 90 seconds and radius of $1.5\arcmin$.
This scan pattern ensures that each detector passes over the point source (within the primary beam)
multiple times in a scan. Additionally, a radius of $1.5\arcmin$ ensures adequate sampling of the
error beam.

Table \ref{table:integrations} shows the time spent (on source) observing each cluster.
MACS 0647, which had exceptionally good quality data compared to the other cluster, 
fares better than the expected $t^{-1/2}$ scaling relative to Abell 1835. 
In Table \ref{table:integrations} the noise is calculated from
a noise map, produced as described in Section \ref{sec:mustangredux}. Specifically, we smooth the map by
convolving it with a Gaussian with FWHM of $10\asec$
and then select the central arcminute ($r < 1\arcmin$), where the weight is nearly uniform, and calculate 
the root-mean-square (RMS) of the selected pixels.

\begin{deluxetable}{c|ccc}
\tabletypesize{\footnotesize}
\tablecolumns{5}
\tablewidth{0pt} 
\tablecaption{Dataset Properties for MUSTANG observations \label{table:integrations}}
\tablehead{
Cluster & Time  & Secondary  & Noise \\
        & (hrs) & Calibrator & ($\mu$Jy/beam) 
}
\startdata
    A1835        & 8.6  & 1337-1257 & 53.3 \\
    MACS J0647.7 & 16.4 & 0721+7120 & 20.8 
\enddata
\end{deluxetable}

\subsection{MUSTANG Beam}
\label{sec:mustangbeam}

The MUSTANG beam is characterized by making calibrated maps of compact sources.
Point source maps of bright secondary calibrators, which  are in a high signal-to-noise regime,
are produced with gentler filtering than is 
used to create galaxy cluster maps(see Section \ref{sec:mustangredux});
the primary difference is that no common mode is subtracted
and we use data from outside the central arcminute to calculate and subtract our polynomial fit.
For a single map, the point source centroid is determined by fitting a two dimensional Gaussian to the
map using a Levenberg-Marquardt least squares minimization in IDL \citep[MPFIT, ][]{markwardt2009}. 
With the centroid determined, pixel values and weights are stored as a function of radius, and the
values are normalized based on the peak of the fitted 2D Gaussian. This constitutes the normalized
point source profile, or MUSTANG beam profile, for one map. 2D Gaussians are fit to all 185 secondary 
calibration observations from fall 2011 until spring 2013 with fair quality data: wind speeds below 
5 m/s, and FWHM $< 11\asec$. The FWHM is taken as the geometric mean of the FWHM along the two axes 
as fit for below. 

\begin{figure}[h]
\begin{center}
	\includegraphics[width=0.5\textwidth]{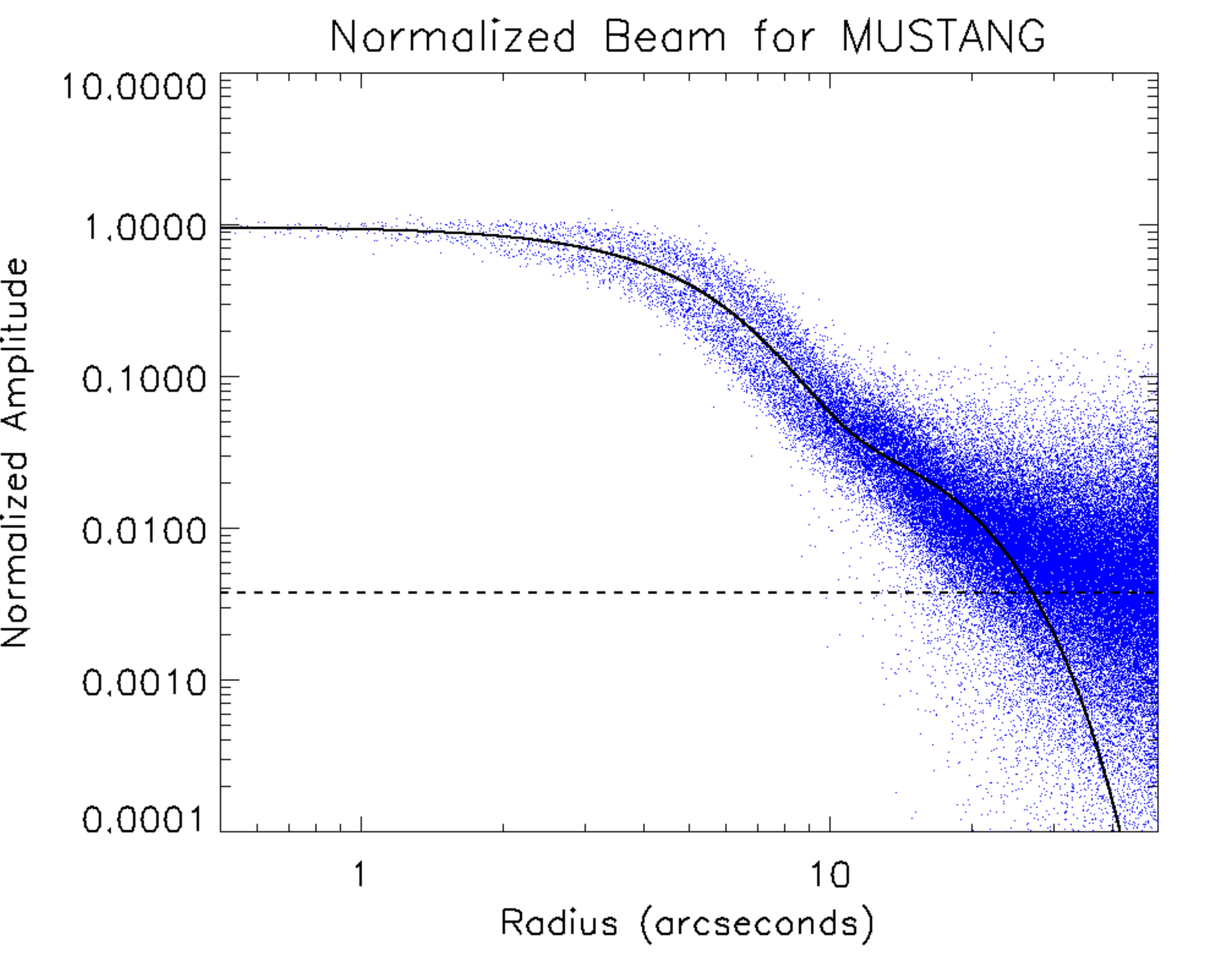}
\end{center}
  \caption{The solid line shows the fitted radial beam profile of MUSTANG. Fitting a double Gaussian, the primary beam has a FWHM of $8.7\asec$, 
    with normalization, $B_1$, of $0.94$. The secondary beam has a FWHM of $28.4\asec$ with a normalization $B_2$, of $0.06$.
    The dashed line shows the weighted RMS of normalized pixels beyond $27\asec$; it intersects the fitted line at $27\asec$.}
  \label{fig:MUSTANG_beam}
\end{figure}

To characterize the typical beam for MUSTANG, we assume a one dimensional, double Gaussian:
\begin{equation}
  B(r) = B_1 e^{-\frac{r^2}{2 \sigma_1^2}} + B_2 e^{-\frac{r^2}{2 \sigma_2^2}}
\end{equation}
where $B_1$ is the normalization of the primary beam, and $B_2$ is the normalization of the secondary (error)
beam. This model is then fit, using MPFIT, to the normalized, azimuthal profiles of all the secondary 
calibrators to obtain $B_1 = 0.94_{-0.02}^{+0.02}$, $\sigma_1 =3.69_{-0.14}^{+0.23}$\asec, $B_2 = 0.06_{-0.02}^{+0.02}$, 
$\sigma_2 = 12.1_{-2.8}^{+3.3}\asec$. This 
corresponds to a primary beam with FWHM of $8.7\asec$ and a secondary (error) beam with FWHM of $28.4\asec$.
Figure \ref{fig:MUSTANG_beam} shows the fitted beam and normalized pixel
values.\footnote{This is also well approximated as a single Gaussian of FWHM $9\asec$.}
The secondary beam is qualitatively consistent with the expected near-sidelobes on the GBT given the MUSTANG
illumination pattern and medium-scale aperture phase errors not fully corrected by the OOF procedure
in Section \ref{sec:musobs}. 

Given the dispersion in pixel values in Figure \ref{fig:MUSTANG_beam}, we investigate the stability
and uncertainty in the fitted parameters. To investigate the stability of the fitted MUSTANG beam,
we divide the sample by observing season (fall to spring). Fitting a one dimensional, double Gaussian to
the beam profile over five seasons of data, we find minimal variation and consistent dispersion in pixel values. Moreover,
the apparent bimodality in the inner beam is present in all seasons, and appears to be due the ellipticity
of individual beams. From the original two-dimensional (single Gaussian) fits, we find the mode and median of 
major-to-minor axis ratios are 1.05 and 1.08 respectively. Taking azimuthal profiles of the fitted (model)
two-dimensional Gaussians reveals the same bimodality.

\subsection{MUSTANG Reduction}
\label{sec:mustangredux}

Processing of MUSTANG data is performed using a custom IDL pipeline. Raw data is recorded as time ordered data (TOD)
from each of the 64 detectors. An outline of the data processing for each scan on a galaxy cluster is given below.
  
  (1) We define a pixel mask from the nearest preceding CAL scan; unresponsive detectors are masked out.
  The CAL scan provides us with unique gains to be applied to each of the responsive detectors.

  (2) A common mode template is calculated as the arithmetic mean of the TOD across detectors. The pulse tube used to cool 
  the array
  produces a coherent 1.411 Hz signal across all detectors. A sinusoid is used as a template to fit this signal.
  The common mode template, pulse tube template, and a polynomial of order $N$ are then simultaneously fit to each detector
  and then subtracted. The polynomial order is given by $N = t_{scan}/t_{poly}$, where $t_{scan}$ is the scan
  duration, $t_{poly} = 2$\amin$ / \bar{v}$, and $\bar{v}$ is the mean scan speed. Typically, $t_{poly}$ is roughly $1.0$
  seconds, while scan durations vary between $90$ and $180$ seconds.
  Subtracting the common mode is powerful at removing atmospheric emission, but has the downside of 
  removing astronomical signals much larger than the instrument FOV. 
  For Abell 1835 ($z=0.25$), $42\asec$ corresponds to 166 kpc; for MACS 0647 ($z=0.59$), $42\asec$ corresponds to 285 kpc.

  (3) After the common mode and polynomial subtraction, each scan undergoes further data quality checks: 
  spike (glitch) rejection, skewness, and Allan variance. Spikes are flagged such that the remaining TOD is still used, 
  while detectors that fail skewness and Allan variance checks are masked for that scan. In total, MACS 0647 has
  11\% of its data flagged, and Abell 1835 has 36\% of its data flagged.
 
  (4) Individual detector weights are calculated as $1/ \sigma_i^2$, where $\sigma_i$ is the RMS of the non-flagged
  TOD for that detector. 

  (5) Maps are produced by gridding the TOD in $1\asec$ pixels in Right Ascension (R.A.) and Declination (Dec). A weight map is
  produced in addition to the signal map. Unsmoothed signal-to-noise (SNR) maps are produced by dividing the signal map by the
  inverse square root of the weight map. For smoothed SNR maps, the signal and variance maps are smoothed,
  and the SNR map is then calculated as the signal map divided by the square root of the variance map.

\subsection{MUSTANG Noise}
\label{sec:mustangnoise}

Because we are in the small signal limit, we need to understand our noise very well in both the time domain and
map (spatial) domain. In the map domain, we can produce noise maps by sending our TOD through the above reduction process
and either flipping, i.e. reversing, the TOD per scan in the time domain, or by flipping sign of the gain between scans. 
The former works by
no longer allowing the signal to be coherently matched to location on the sky, while the latter works by effectively
cancelling any signal observed. Thorough analysis has shown better behavior in the gain-flipped maps. For instance,
if we make an unsmoothed SNR map from the gain-flipped TOD, we find a mean of 0 with a standard deviation of 1.
In both the gain- and time-flipped noise maps, we find that pixel weights accurately reflect the RMS
of pixel values within a weight bin, and the pixel values follow a Gaussian distribution.
Smoothed SNR maps produced from either of the flipped TOD methods produce standard deviations greater than 1; the
advantage of the gain-flipped smoothed SNR maps is that their means are 0, whereas the time-flipped SNR maps have 
means that are offset from 0. The standard deviation of our smoothed noise SNR maps ($\sigma_{SNR}$) tending towards values 
greater than 1 indicates our smoothing procedure does not accurately handle the weights of each pixel; 
thus, we use $\sigma_{SNR}$ to correct our true (not noise) SNR maps. For our canonical smoothing kernel ($10\asec$
FWHM), $\sigma_{SNR} \sim 1.6$; we then divide the true SNR map by this factor. As the model fitting presented in this work
makes use of only the non-smoothed maps, this correction factor is only used for visualizing smoothed SNR maps.
 

\subsection{Bolocam Observations and Reduction}
\label{sec:bolocamredox}

Bolocam is a 144-element camera that was a facility instrument on the Caltech Submillimeter Observatory (CSO) from
2003 until 2012. Its field of view is $8'$ in diameter, and at 140 GHz it has a resolution of $58\asec$ FWHM
(\citet{glenn1998,haig2004}). The clusters were observed with a Lissajous pattern that results in a tapered
coverage dropping to 50\% of the peak value at a radius of roughly $5'$, and to 0 at a radius of $10'$.
The Bolocam maps used in this analysis are $14\arcmin \times 14\arcmin$.

The Bolocam data are the same as those used in \citet{czakon2014} and \citet{sayers2013};
the details of the reduction are given therein, along with \citet{sayers2011}. Bolocam observed Abell
1835 for 14.0 hours resulting in a noise of 16.2 $\mu K_{CMB}$-arcminute, and observed MACS 0647 for 11.4 hours
resulting in a noise of 22.0 $\mu K_{CMB}$-arcminute.
Overall, the reduction and calibration is quite similar to that used for MUSTANG, and Bolocam achieves a 
5\% calibration accuracy and $5\asec$ pointing accuracy.


\section{Joint Map Fitting Technique}
\label{sec:jointfitting}


\subsection{Producing Cluster Model Maps}
\label{sec:makemodelmaps}
We choose to construct 3D electron pressure profiles as parameterized by a generalized Navarro, Frenk,
and White profile \citep[hereafter, gNFW][]{navarro1997,nagai2007}:
\begin{equation}
  \Tilde{P} = \frac{P_0}{(C_{500} X)^{\gamma} [1 + (C_{500} X)^{\alpha}]^{(\beta - \gamma)/\alpha}}
\end{equation}
where $X = R / R_{500}$, and $C_{500}$ is the concentration parameter; one can also write ($C_{500} X$) as
($R / R_s$), where $R_s = R_{500}/C_{500}$. $\Tilde{P}$ is the electron pressure in units of the characteristic
pressure $P_{500}$.
The 3D pressure profile is assumed to be spherical and is integrated along the line of sight to produce 
a Compton $y$ profile, given as 
\begin{equation}
  y(r) = \frac{P_{500} \sigma_{T}}{m_e c^2} \int_{-\infty}^{\infty} \Tilde{P}(r,l) dl
\end{equation}
where $R^2 = r^2 + l^2$, $r$ is the projected radius, and $l$ is the distance from the center of the cluster
along the line of sight. Once integrated, $y(r)$ is gridded as $y(\theta)$ and  is realized as a map 
(pixels of $1\asec$ and $20'$ on a side) using the \emph{Archive of Chandra Cluster Entropy Profile Tables} 
(ACCEPT, \citet{cavagnolo2009}) centroid for the
cluster. From here, we produce two model maps: one for Bolocam and one for MUSTANG. In each case,
we convolve the Compton $y$ map by the appropriate beam shape. For Bolocam we use a Gaussian with FWHM
$= 58\asec$, and for MUSTANG we use the double Gaussian as found in Section \ref{sec:mustangbeam}.

The convolved maps are then regridded to the same scale and map size as the reduced data maps for each
instrument. The regridded Bolocam model map is then convolved with a 2D filter function that describes
the effects of Bolocam data processing \citep{sayers2011}. The MUSTANG 
map is filtered by converting the model map into model TOD, using the true TOD from a galaxy cluster 
as a template (namely for telescope pointing trajectory). The model TOD can then be processed using
the same custom IDL pipeline used to reduce the data to create the filtered MUSTANG model map.


\subsection{Point Source Model maps}
\label{sec:ptsrc_models}

While we do not detect a point source in MACS 0647 (either in Bolocam or MUSTANG), we clearly detect a
point source in Abell 1835 in the MUSTANG image. For the Bolocam image, the point source in Abell 1835
has been subtracted based on an extrapolation of a power law fit to the 1.4 GHz NVSS \citep{condon1998}
and 30 GHz SZA \citep{mroczkowski2009} measurements, leading to an assumed flux density of 
$0.77 \pm 0.24$ mJy \citep{sayers2013}.

For MUSTANG, where a point source is detected at high significance in our galaxy cluster map,
we take the following approach:
\begin{enumerate}
  \item For each cluster, the same process as in Section \ref{sec:mustangbeam} is applied to only those 
    secondary calibrators which were observed during the same sessions as that particular cluster.
  \item The fitted profile is then evaluated as a map, with the centroid and total amplitude as
    determined by the 2D Gaussian fit, and shape as determined by the 1D fit.
  \item The point source map is filtered in the same manner in which the cluster model map is filtered,
    and the resultant image is used as our point source component in model fitting, with the normalization
    as a free parameter in the fit.
\end{enumerate}


\subsection{Parameter Space Searched}
\label{sec:param_space}

Given that the spatial coverage from MUSTANG and Bolocam is well suited to constraining the inner pressure profile,
we choose to allow the gNFW parameters $\gamma$, $C_{500}$, and $P_0$ to vary. To reduce
degeneracies, we fix $\alpha$ and $\beta$.
We choose our fixed parameters from four established 
sets of gNFW parameters: those found in \citet[][hereafter N07]{nagai2007},
\citet[][hereafter A10]{arnaud2010}, \citet[][hereafter P12]{planck2013a}, and \citet[][hereafter S13]{sayers2013}; 
these are summarized in Table \ref{table:gnfw_params}. 
We construct a model map for each set of $\alpha$, $\beta$, $\gamma$, and $C_{500}$, and assume a starting
value for $P_0$, which is then determined in our fits.
Since $\alpha$ and $\beta$ are fixed,
each cluster is initially searched over $0 < \gamma < 1$ in steps of $\delta \gamma = 0.1$, and over
$0.1 < C_{500} < 2.1$ in steps of $\delta C_{500} = 0.1$. These ranges are refined after the first pass, 
and generally $\delta \gamma$ is reduced to 0.05. To create models in finer steps than $\delta \gamma$ 
and $\delta C_{500}$, we interpolate filtered model maps from nearest neighbors from the grid of original 
filtered models.

\begin{deluxetable}{c | c c c c c}
\tabletypesize{\footnotesize}
\tablecolumns{5}
\tablewidth{0pt} 
\tablecaption{Parameters of gNFW models considered \label{table:gnfw_params}}
\tablehead{
    Model & $C_{500}$  & $\alpha$ & $\beta$ & $\gamma$ & $P_0$ 
}    
\startdata
    N07 & 1.80 & 1.30  & 4.30 & 0.71 & 3.94 \\
    A10 & 1.18 & 1.05  & 5.49 & 0.31 & 7.82 \\
    P12 & 1.81 & 1.33  & 4.13 & 0.31 & 6.54 \\
    S13 & 1.18 & 0.86  & 3.67 & 0.67 & 4.29 
\enddata
\tablecomments{We considered these four sets of models and fix $\alpha$ and $\beta$ for each.}
\end{deluxetable}

Given that MUSTANG is sensitive to substructure, we fix the MUSTANG cluster model centroid to the position of
 the ACCEPT centroid. If the MUSTANG centroid is allowed to vary, we find that the fit can significantly be
influenced by the cluster substructure. As a result, such fits generally do not accurately represent the
bulk cluster component that we seek to model.
However, Bolocam maps are dominated by the bulk SZE signal from the cluster, and have pointing accuracy
of 5\asec. Thus we allow the Bolocam 
pointing is allowed to vary up to a total range of $10\asec$ in R.A. and Dec relative to the ACCEPT
centroid with a Gaussian prior of $\sigma_{cen} = 5\asec$.


\subsection{Least Squares Fitting}
\label{sec:lsq_fitting}

A set, spanning $\gamma$ and $C_{500}$, of model maps for each of the established parameter sets 
(Section \ref{sec:param_space}) is created for each galaxy cluster.  From each model map, we construct 
a model array of pixel values, $\overrightarrow{d}_m$, from a subset of the pixels in the model maps to fit 
to the data array of pixel values, $\overrightarrow{d}$, selected from the same subset of pixels in the data map.
The subset is chosen to exclude pixels with low coverage; principally we select the inner arcminute of the 
MUSTANG maps, and a $14$\amin $\times 14$\amin box for Bolocam data.

This requires that the data map and model map have the exact same astrometry and pixelization.
These two arrays can be constructed from the model and data for (1) MUSTANG, (2) Bolocam, or (3) 
MUSTANG and Bolocam. In the third case, the arrays are the concatenation of the two arrays in the
first and second cases.
We then assume that we can construct our model map as a linear combination of components (e.g. a bulk pressure
profile and in the case of Abell 1835, a point source). To do this, we construct an $N \times M$ matrix, $\mathbf{A}$, where $N$ is the
number of data points in our data array, $\overrightarrow{d}$, and $M$ is the number of model components used.
This can be written as:
\begin{equation}
  \overrightarrow{d}_m = \mathbf{A} \overrightarrow{a}_m.
\end{equation}
We can then use the $\chi^2$ statistic as our goodness of fit
\begin{equation}
  \chi^2 = (\overrightarrow{d} - \overrightarrow{d}_m)^T \mathbf{N}^{-1} (\overrightarrow{d} - \overrightarrow{d}_m),
\end{equation}
and find that the minimum $\chi^2$ is achieved when:
\begin{equation}
  \overrightarrow{a}_m = (\mathbf{A}^T \mathbf{N}^{-1} \mathbf{A})^{-1} \mathbf{A}^T \mathbf{N}^{-1} \overrightarrow{d}.
  \label{eqn:comp_amps}
\end{equation}
Here, $\mathbf{N}$ is the covariance matrix, which is formally defined as:
\begin{equation}
  \mathbf{N}_{ij} = <d_i d_j> - <d_i><d_j>
\end{equation}
Here, $d_i$ is the pixel value in a noise realization map, and the average is taken for a given pixel of
several noise realization maps.

In practice, we take $\mathbf{N}_{ij}$ to be a diagonal matrix for MUSTANG and Bolocam, 
with $\mathbf{N}_{ij} = \delta_{ij}/w_i$, where
$\delta_{ij}$ is the Kronecker delta, and $w_i$ is the weight for pixel $i$.
MUSTANG's detector noise is dominated
by phonon noise; thus with the common mode (which is dominated by atmospheric noise) subtracted,
we expect the map pixel noise to be uncorrelated. Indeed, taking jackknife resampling of MUSTANG
data has been used to produce a correlation matrices for clusters and verifies that assuming 
uncorrelated pixel noise in MUSTANG maps is valid.

\begin{figure}[!ht]
  \begin{center}
  \includegraphics[width=0.5\textwidth]{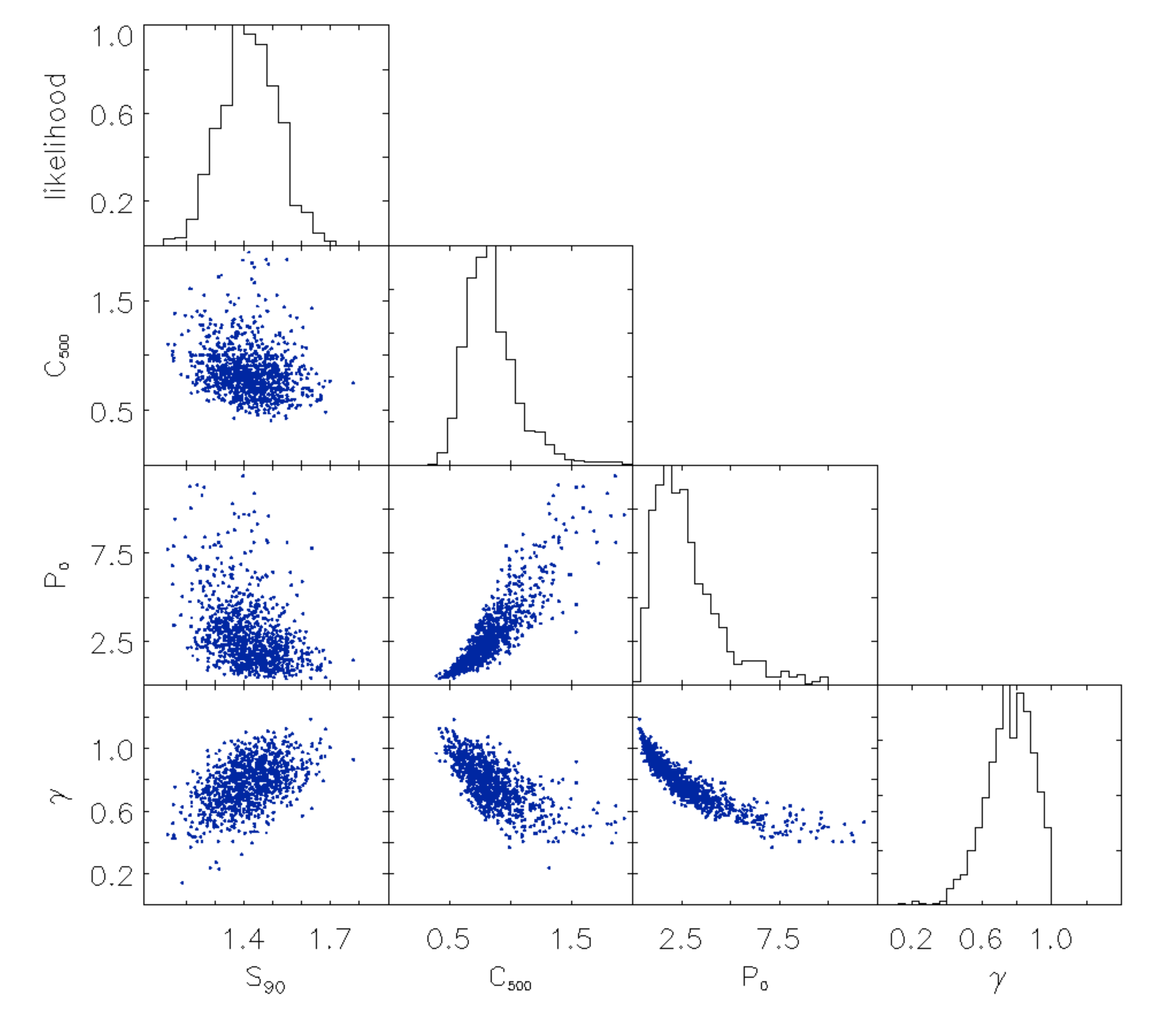}
  \end{center}
  \caption{Results from bootstrap Monte Carlo fits to 1000 realizations for Abell 1835 with A10 values for 
    $\alpha$ and $\beta$. Here we can determine the scatter in each of the fitted gNFW parameters. $S_{90}$ 
    is the flux density at 90 GHz as determined by MUSTANG.}
  \label{fig:a1835_density_plot}
\end{figure}


If we simply append the MUSTANG model array to the Bolocam model array:
\begin{equation}
  \overrightarrow{d}_m = [ \overrightarrow{d}_{m,Bolocam}, \overrightarrow{d}_{m,MUSTANG} ]
\end{equation}
then solving this set of equations would be a linear problem. However, we add a calibration offset term, 
$k$, to allow for offsets in calibration between MUSTANG and Bolocam data, and create the following array:
\begin{equation}
  \overrightarrow{d}_{m,cal} = [ \overrightarrow{d}_{m,Bolocam} ,  k * \overrightarrow{d}_{m,MUSTANG} ]
  \label{eqn:data_w_cal}
\end{equation}
We cast Equation \ref{eqn:data_w_cal} as such because we expect the normalization
given by MUSTANG and Bolocam to be equal, except for any calibration offset between instruments.
Furthermore, we can quantify the calibration uncertainties and thus put a prior on it.
Solving Equation \ref{eqn:comp_amps} is no longer a problem of linearly independent variables.
Thus, we use MPFIT to quickly solve for $\overrightarrow{a}_m$ and obtain a pseudo-$\chi^2$ ($\Tilde{\chi}^2$).
To calculate $\Tilde{\chi}^2$, we use the same formulation as above, but redefine
\begin{align*} 
  \overrightarrow{d}_{cal}   &= [ \overrightarrow{d}_{old} , 0.0 ] \\
  \overrightarrow{d}_{m,cal} &= [ \overrightarrow{d}_{m,old} , k ] = \mathbf{A}_{new} \overrightarrow{a}_{m,cal} \\
\end{align*}
where $\overrightarrow{a}_{m,cal} = [\overrightarrow{a}_m, k]$, and expand $\mathbf{N}$ to allow for the extra fitted
value.  $\mathbf{N}$ is modified simply by assigning zero for all off-diagonal terms, 
and adding the expected variance of $k$ on the diagonal term. We adopt a calibration uncertainty of 11.2\%, 
which accounts for the calibration accuracies of Bolocam and MUSTANG cited in Section \ref{sec:musobs}. 


\subsection{Uncertainties}
\label{sec:uncertainties}

\begin{figure}[!h]
\begin{center}  
	\includegraphics[width=0.5\textwidth]{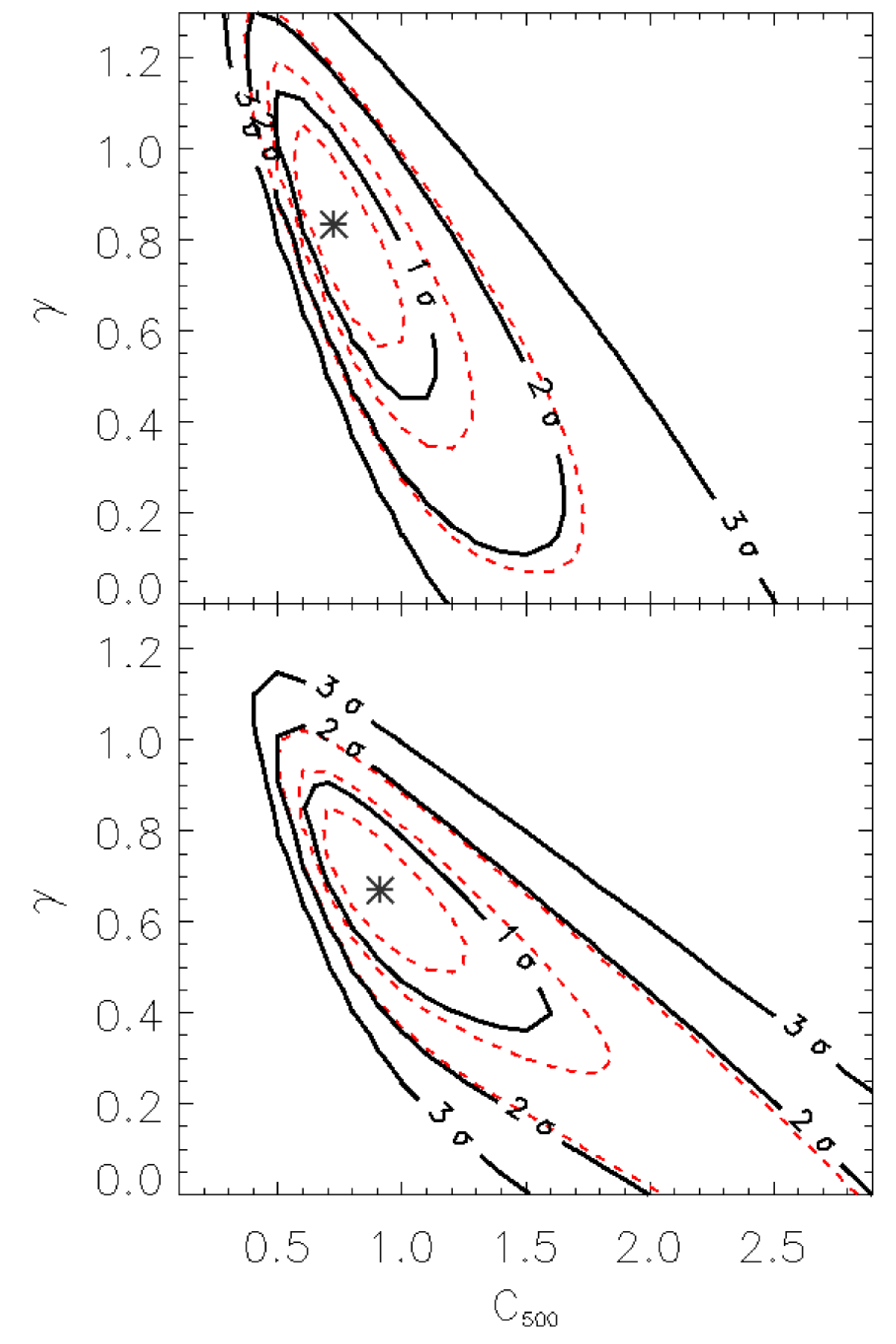}
\end{center}
 \caption{Confidence intervals for Abell 1835 for A10 values of $\alpha$ and $\beta$. 
    Top panel shows the confidence intervals for Bolocam only fit.
    Bottom panel shows the confidence intervals for the joint fit. Dashed lines show confidence intervals based on
    $\Delta \Tilde{\chi}^2$. Solid lines are the true confidence intervals based on the bootstrap MC
    results. The asterisks denote the best fit.}
  \label{fig:a1835_ci_params}
\end{figure}

For each set of fixed $\alpha$ and $\beta$ (see Table \ref{table:gnfw_params}),
the joint fit procedure returns a $\Tilde{\chi}^2$. This is used to plot contours of
$\Delta \Tilde{\chi}^2$, and the equivalent confidence intervals. Following \citet{sayers2011}, the 
uncertainty for each parameter is estimated via the 1$\sigma$ confidence interval of the best fits 
over 1000 noise realizations added to model clusters. Furthermore, because $k$ is not an independent parameter, 
$\Tilde{\chi}^2$ will not provide a fully accurate assessment of the confidence intervals. 

We create 1000 realizations about the best fit model by adding 1000 instances of noise (for both Bolocam
and MUSTANG) to the best fit model. Then, we find the best fit to each realization. The results of these
fits is shown in Figure \ref{fig:a1835_density_plot}. The noise realizations for
Bolocam are precomputed from previous work \citep{sayers2013}. Noise realizations for MUSTANG are computed by 
random number generation with Gaussian distribution
for each pixel, with $\sigma_i = \sqrt{1/w_i}$, based on the same weight used to calculate the noise matrix.
These fits are tabulated and used to rescale the $\Delta \Tilde{\chi}^2$ confidence intervals as seen in
Figure \ref{fig:a1835_ci_params}. The rescaling of the confidence intervals is primarily due to the
non-diagonality of the noise in the Bolocam covariance matrix.

We also investigated the potential impact from the uncertainty in the point source. For a given cluster,
we follow the steps in Section \ref{sec:ptsrc_models}, and we then calculate point source uncertainty 
models which adopt the $1\sigma$ values for the width of the main beam, as reported in Section \ref{sec:mustangbeam},
and fit the remaining components. The fitting procedure is then rerun twice: once with each of these models. 
Neither the fitted gNFW parameters nor the confidence intervals change. However, across the three point source models
(two uncertainty and best-fit point source models), the minimum $\Tilde{\chi}^2$ does change in the expected manner: 
it is greater for both of the uncertainty models than the best fit point source model.


\section{Results}
\label{sec:results}

\subsection{Abell 1835 (z=0.25)}

\begin{figure}[!ht]
\begin{center} 
  \includegraphics[width=0.5\textwidth]{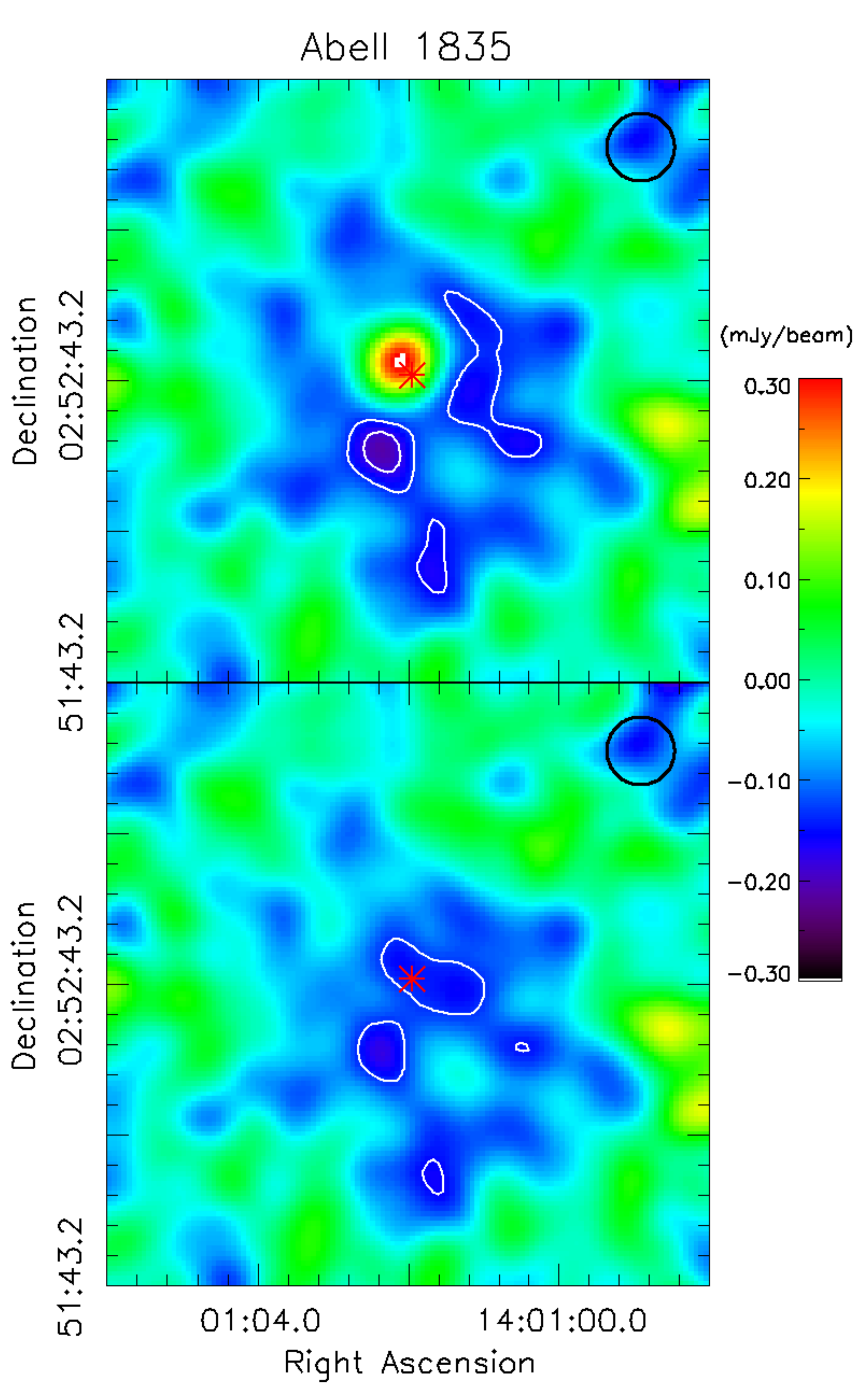}
\end{center}
\caption{MUSTANG SZE specific flux density map of Abell 1835. The white contours are at the 3 and 4 $\sigma$ significance
    level. The ACCEPT centroid is the red asterisk. The positive region close to the asterisk is the point source.
    The black circle in the upper right represents the 
    effective beam of FWHM of $13.5\asec$ (i.e. our $9\asec$ telescope beam, smoothed by a $10\asec$ FWHM Gaussian.)}
  \label{fig:a1835_map}
\end{figure}

Abell 1835 is a well studied massive cool core cluster. The cool core was noted to have substructure in the central
$10\asec$ by \citet{schmidt2001}, and identified as being due the central AGN by \citet{mcnamara2006}. Abell 1835 has also
been extensively studied via the SZE \citep{reese2002,benson2004,bonamente2006,sayers2011,mauskopf2012}. The models adopted
were either beta models or generalized beta models, and tend to suggest a shallow slope for the pressure interior
to $10\asec$. Previous analysis of Abell 1835 with MUSTANG data \citep{korngut2011} detected the SZE decrement, but not
at high significance, which is consistent with a featureless, smooth, broad signal. Our updated MUSTANG reduction
of Abell 1835 is shown in Figure \ref{fig:a1835_map}, and shows the same features as in \citet{korngut2011}.

\begin{deluxetable}{c|ccccccc}
\tabletypesize{\footnotesize}
\tablecolumns{5}
\tablewidth{0pt} 
\tablecaption{Best fit gNFW parameters for Abell 1835 \label{table:a1835_gnfw_fits}}
\tablehead{
    Model & $C_{500}$  & $\alpha$ & $\beta$ & $\gamma$ & $P_0$ & k & $\Tilde{\chi}^2$
}
\startdata
    N07 & $1.44_{-0.22}^{+0.71}$ & 1.30  & 4.30   & $0.74_{-0.13}^{+0.15}$  & $3.43_{-1.78}^{+1.41}$  & 1.18 & 12837 \\
    A10 & $0.83_{-0.15}^{+0.35}$ & 1.05  & 5.49   & $0.75_{-0.17}^{+0.14}$  & $2.54_{-1.37}^{+1.25}$  & 1.15 & 12835 \\
    P12 & $1.45_{-0.28}^{+0.35}$ & 1.33  & 4.13   & $0.84_{-0.12}^{+0.18}$  & $2.80_{-1.15}^{+1.54}$  & 1.14 & 12838 \\
    S13 & $2.29_{-0.52}^{+1.30}$ & 0.86  & 3.67   & $0.36_{-0.21}^{+0.33}$  & $19.3_{-6.16}^{+9.75}$  & 1.19 & 12831
\enddata
\tablecomments{$\gamma$, $P_0$, $C_{500}$, and k, the calibration offset, were varied.
    The degrees of freedom were 12880.}
\end{deluxetable}

We find the best fit A10 model (all parameters but $P_0$ fixed to A10 values) has $\Tilde{\chi}^2 = 12861$. To calculate
$\Tilde{\chi}^2$ for no cluster model, we fix the point source amplitude to that found in the previous fit. We find
$\Delta \Tilde{\chi}^2 = 892.1$, with $\Delta$DOF$= 2$ corresponds to $29.7 \sigma$ significance. For Bolocam only, the $\Delta
\chi^2$ between a cluster model being fit or not yields a $28.9 \sigma$ detection, while the for MUSTANG we find a
$10.2 \sigma$ detection of an A10 model from $\Delta \chi^2$.

\begin{figure}[!h]
\begin{center} 
  \includegraphics[width=0.5\textwidth]{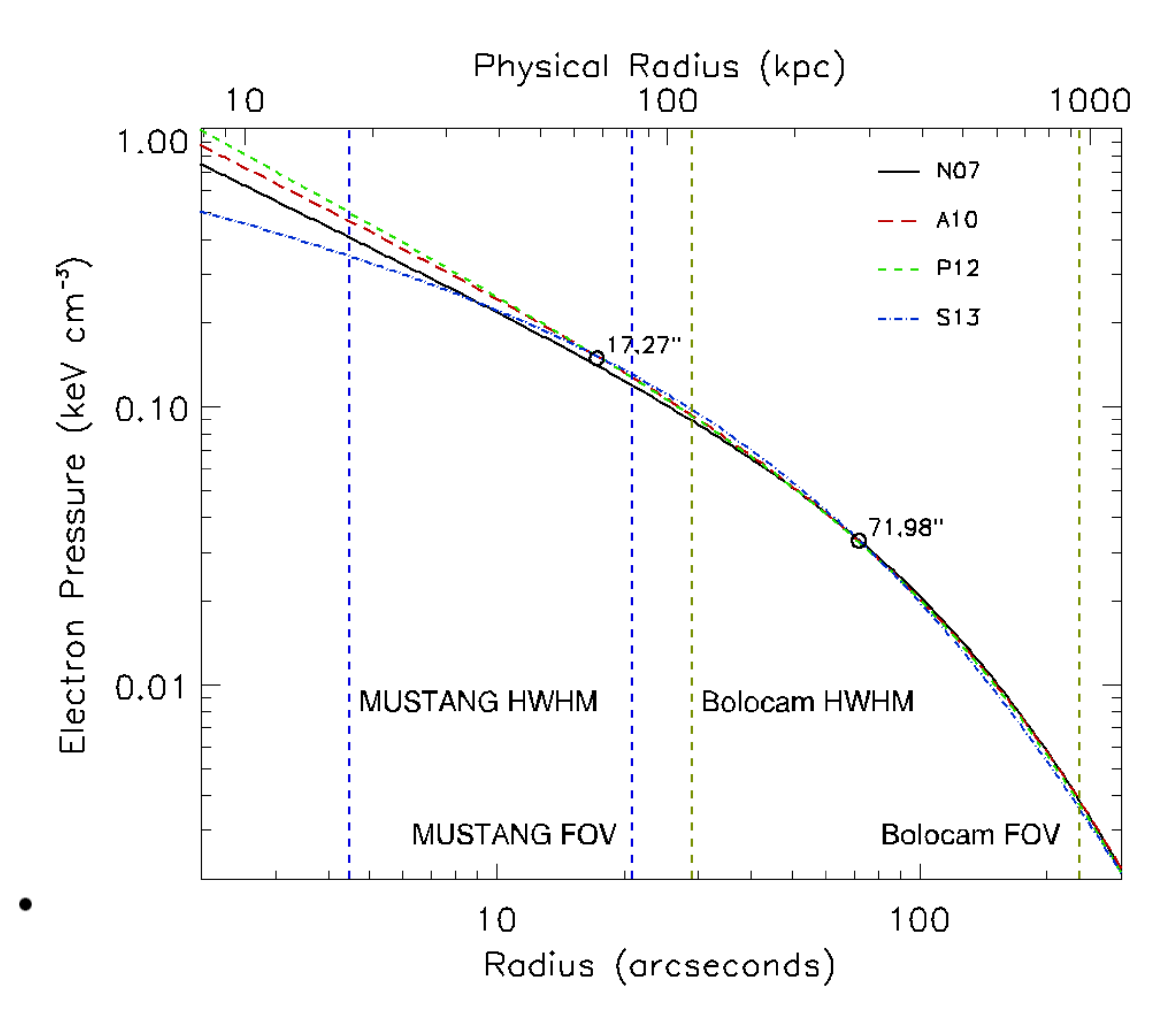}
\end{center} 
  \caption{Best fit pressure profiles for Abell 1835 for the different sets of fixed $\alpha$ and $\beta$, denoted by the model
   it is taken from. The circles denote radial ranges where the pressure profiles show (local) minimum scatter.}
  \label{fig:a1835_dpsets}
\end{figure}

Our best joint fit over the four sets of $\alpha$ and $\beta$, shown in Table \ref{table:a1835_gnfw_fits}, comes from
the S13 values of $\alpha$ and $\beta$: the best fit floated parameters are: $\gamma = 0.36$, $P_0 = 19.3$, and $C_{500}=2.28$.
Despite variations in the best fit $\gamma$ values, Figure \ref{fig:a1835_dpsets} shows the best joint fit pressure profiles
of Abell 1835 for each of the four sets of fixed $\alpha$ and $\beta$, are in good agreement with each other. 
Moreover, we find that the four model fits achieve minimum scatter at two separate radii, roughly corresponding to the
geometric mean between the resolution and FOV of each instrument.

We find the point source in the MUSTANG map at R.A.=14:01:02.1, Dec=02:52:47 is best fit with a flux 
density of $1.38 \pm 0.10$ mJy, and has a
correlation coefficient of 0.076 with the cluster amplitude. This minimal degeneracy can also be seen in 
Figure \ref{fig:a1835_density_plot}. Similarly, changing the assumed beam shape as discussed in Section
\ref{sec:uncertainties} has a negligible change on the flux density. The amplitude of the point source 
suggests a slight flattening of the spectral index between $\nu=1.4$ GHz and $\nu = 28.5$ GHz 
(\citet{condon1998},\citet{reese2002}) of $\alpha_{\nu} = 0.89$, to a spectral index of $\alpha_{\nu} = 0.59$ between 
$\nu = 28.5$ GHz and $\nu = 90$ GHz. Such a spectral index is also consistent with \citet{mcnamara2014}, which find a
spectral index of $\alpha_{\nu} = 0.54$ between $\nu = 92$ GHz and $\nu = 276$ GHz. The assumed flux density of
the subtracted point source for Bolocam, $0.77 \pm 0.24$ mJy at 140 GHz, is consistent with the other measurements.

The point source flux density found with MUSTANG is consistent with that obtained from observations with
the Atacama Large Millimeter/Sub-millimeter Array (ALMA) in
\citet{mcnamara2014}, which find the central continuum source has a flux density of 1.26 $\pm 0.03$ mJy. We note they also
detect a $10^{10} M_{\odot}$ molecular outflow at 92 GHz, with a total integral flux of
$3.6$ Jy km s$^{-1}$ for CO (1-0), which would correspond to an equivalent continuum flux density of 6 $\mu$Jy, 
and would not contribute much additional flux density to the point source flux density as seen by \citet{mcnamara2014}. 
However, this source is reported with a position of R.A.=14:01:02.083, 
Dec=02:52:42.649, which is $4\asec$ offset from the position found in our MUSTANG data. Since we consider our positional
uncertainty to be $2\asec$, this is a larger than typical pointing offset, but is difficult to rule out. A list of selected
point source flux densities is provided in Table \ref{table:a1835_ptsrc}.

Figure \ref{fig:a1835_ci_params} shows that Bolocam, by itself, can place fairly tight constraints on the gNFW model parameters,
primarily on $C_{500}$. That is, for A10 values of $\alpha$ and $\beta$, as in Figure \ref{fig:a1835_ci_params}, Bolocam finds
$C_{500} = 0.73_{-0.17}^{+0.25}$, $\gamma = 0.83_{-0.23}^{+0.22}$, and $P_0 = 1.75_{-1.22}^{+2.51}$, whereas the joint fit yields
$C_{500} = 0.83_{-0.15}^{+0.35}$, $\gamma = 0.75_{-0.17}^{+0.14}$, and $P_0 = 2.54_{-1.37}^{+1.25}$. Across the model sets, the trend
for Abell 1835 is that joint fit tends to loosen the constraint on $C_{500}$, while improving the constraints on $\gamma$ and
$P_0$ relative to the fits to solely Bolocam data.
It is worth noting that for each value of $\gamma$ and $C_{500}$, only $P_0$ is a free parameter in the Bolocam only fits.
In contrast, for the joint fit, the calibration offset and MUSTANG point source amplitude are additional free parameters.
The addition of MUSTANG data slightly reduces the inner slope, $\gamma$, relative to the Bolocam-only fit. This is suggestive
that Bolocam, with subtraction of its adopted point source model, has not underestimated the SZE signal. Moreover, given
the peak decrement of $-20$ mJy in the Bolocam map, an adjustment of $\sim 0.2$ mJy, which is the uncertainty
on the assumed $0.77$ mJy, would negligibly alter the constraints.
Both the Bolocam and joint constraints indicate a relatively steep slope, which is typical for a cool core cluster 
\citep[e.g.][]{arnaud2010,sayers2013}.

\begin{deluxetable}{c c c c c}
\tabletypesize{\footnotesize}
\tablecolumns{5}
\tablewidth{0pt} 
\tablecaption{Abell 1835 point source flux densities \label{table:a1835_ptsrc}}
\tablehead{
    $S_{1.4}^a$ (mJy) & $S_{28.5}^b$ (mJy) & $S_{90}$ (mJy)  & $S_{92}^c$ (mJy) & $S_{276}^c$ (mJy)
}
\startdata
    $41.4\pm 1.9$   & $2.76\pm 0.14$    & $1.38\pm 0.10$ & $1.26\pm 0.03$ & $0.7\pm0.1$
\enddata
  \tablecomments{$S_{90}$ is from this work.}
\tablerefs{$^a$ From \citet{condon1998} $^b$ From \citet{reese2002}
    $^c$ From \citet{mcnamara2014}.}
\end{deluxetable}

\subsection{MACS J0647.7+7015 (z=0.59)}

\begin{figure}[!ht]
\begin{center} 
  \includegraphics[width=0.5\textwidth]{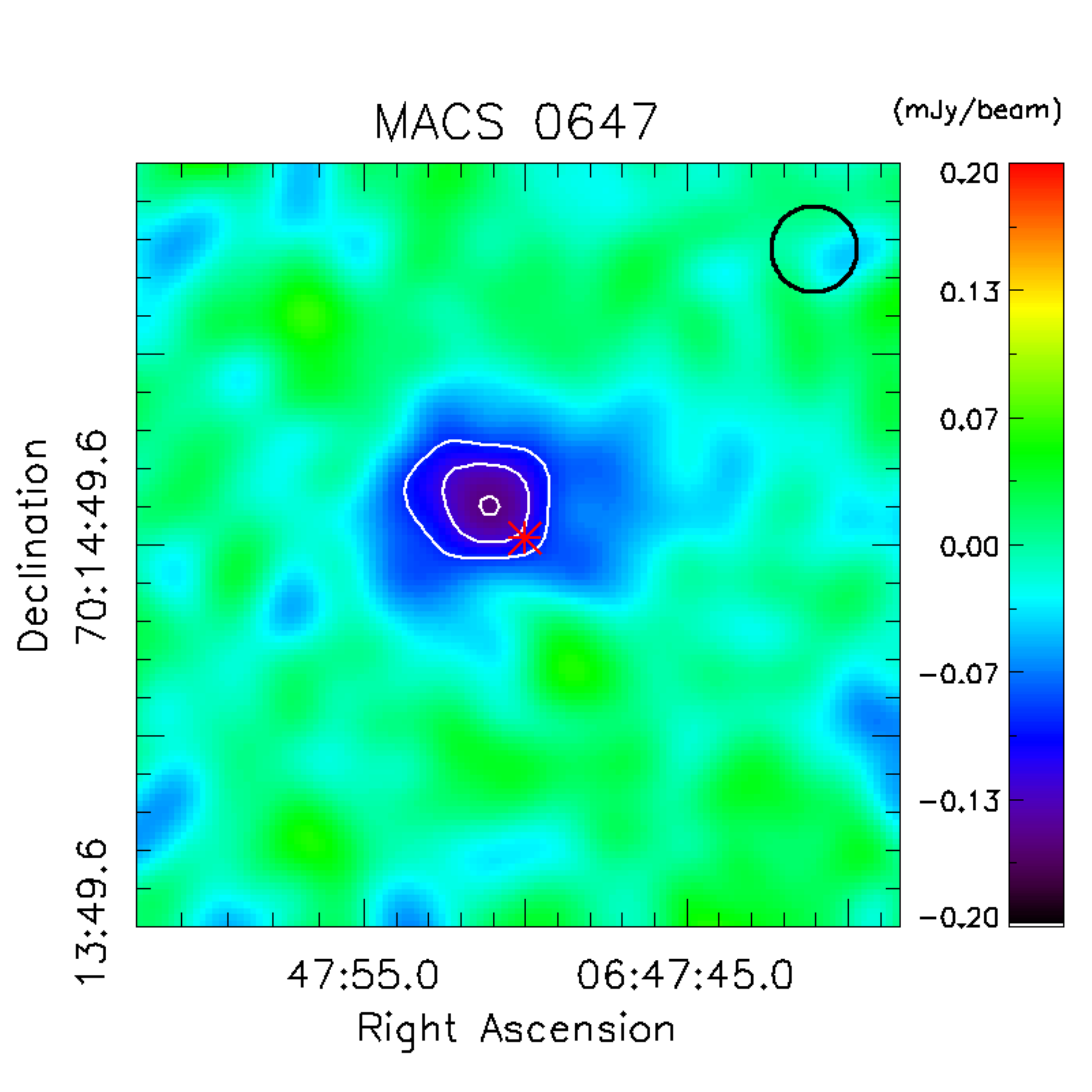}
\end{center} 
  \caption{MUSTANG SZE flux density map of MACS 0647. The white contours are at the 3, 4, and 5 $\sigma$ significance
    level. The ACCEPT centroid is the red asterisk. The black circle in the upper right represents the 
    effective beam of FWHM of $13.5\asec$.}
  \label{fig:m0647_map}
\end{figure}

MACS 0647 is at $z = 0.591$ and is classified as neither a cool core nor a disturbed cluster \citep{sayers2013}. 
It was included in the CLASH sample due to its strong lensing properties \citep{postman2012}.
Gravitational lensing \citep{zitrin2011}, X-ray surface brightness \citep{mann2012}, 
and SZE (MUSTANG, see Figure \ref{fig:m0647_map}, and Bolocam) maps all
show elongation in an east-west direction. 
In the joint analysis presented here, we see that the spherical model provides an adequate fit to both datasets and we note 
that the spherical assumption allows for a easier interpretation of the mass profile of the cluster.

To calculate how significantly Bolocam and MUSTANG detect an SZE bulk decrement, we calculate a $\Delta \Tilde{\chi}^2$ 
as we did for Abell 1835. $\Delta \Tilde{\chi}^2$ is computed as the difference of $\Tilde{\chi}^2$
from an A10 profile (all parameters but $P_0$ fixed to A10 values) fit to our datasets, and $\Tilde{\chi}^2$ assuming
no model. The joint fit (both data sets) yields a 
$26.3 \sigma$ significance, while for Bolocam only, $\Delta \chi^2$ yields a $23.9 \sigma$ detection, and 
in the MUSTANG data we find a $10.8 \sigma$ detection of an A10 model.

\begin{figure}[!h]
\begin{center} 
  \includegraphics[width=0.5\textwidth]{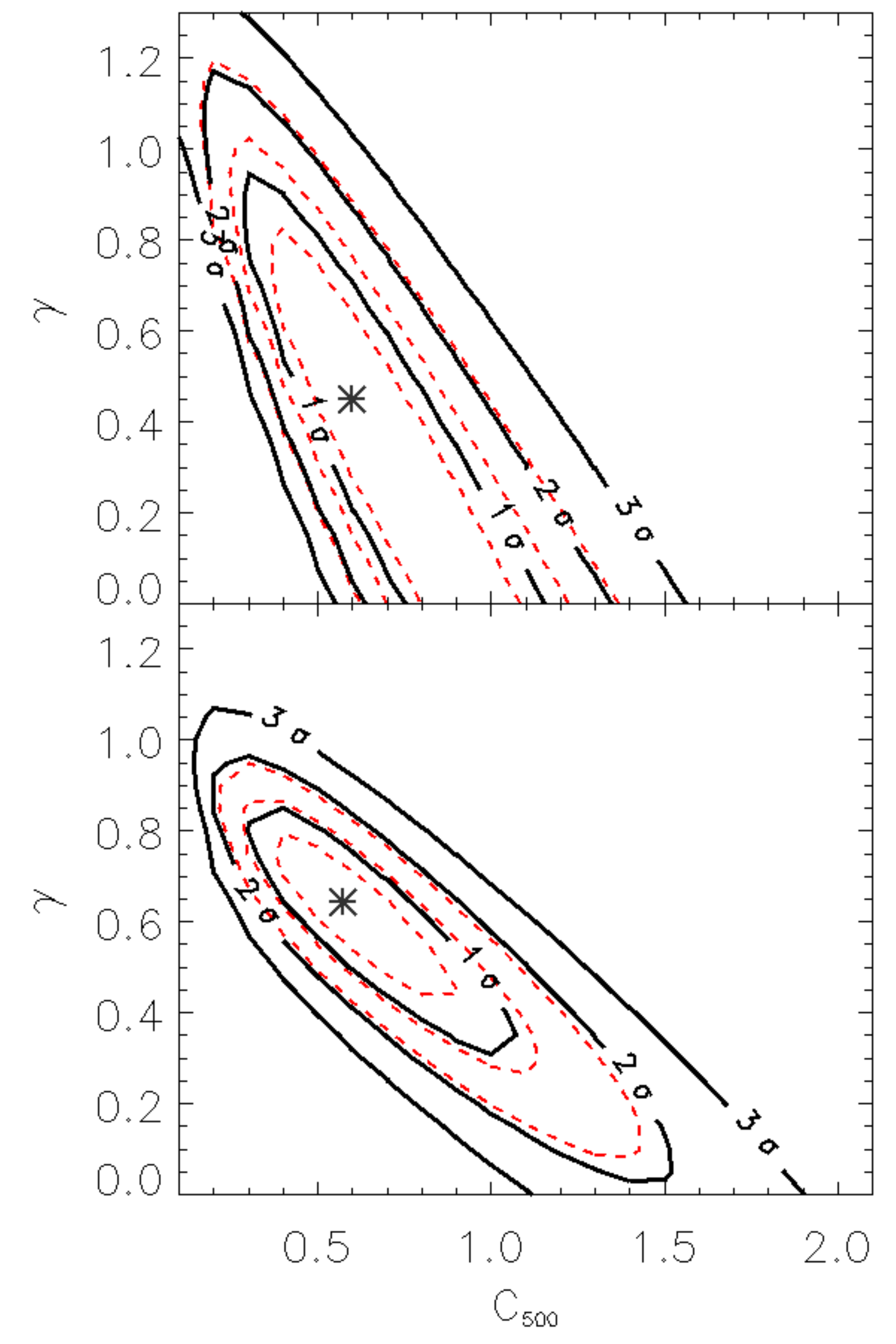}
\end{center} 
  \caption{Confidence intervals for MACS 0647 with A10 values of $\alpha$ and $\beta$. 
    Top panel shows the confidence intervals for Bolocam only fit.
    Bottom panel shows the confidence intervals for the joint fit. Dashed lines show confidence intervals based on
    $\Delta \Tilde{\chi}^2$ at face value. Solid lines are the true confidence intervals based on the bootstrap MC
    results.}
  \label{fig:m0647_ci_params}
\end{figure}

\begin{deluxetable}{c|ccccccc}
\tabletypesize{\footnotesize}
\tablecolumns{5}
\tablewidth{0pt} 
\tablecaption{Best fit gNFW parameters for MACS 0647 \label{table:m0647_gnfw_fits}}
\tablehead{
    Model & $C_{500}$  & $\alpha$ & $\beta$ & $\gamma$ & $P_0$ & $k$ & $\Tilde{\chi^2}$
}
\startdata
    N07 & $0.93_{-0.36}^{+0.31}$ & 1.30  & 4.30  & $0.70_{-0.17}^{+0.10}$ & $2.10_{-1.17}^{+0.93}$ & 1.14  & 12845  \\
    A10 & $0.60_{-0.22}^{+0.25}$  & 1.05  & 5.49 & $0.61_{-0.15}^{+0.12}$ & $2.24_{-1.20}^{+2.03}$  & 1.14  & 12844 \\
    P12 & $1.03_{-0.40}^{+0.32}$  & 1.33  & 4.13 & $0.70_{-0.17}^{+0.10}$ & $2.25_{-1.32}^{+1.04}$  & 1.14  & 12845 \\
    S13 & $1.19_{-0.64}^{+0.54}$  & 0.86  & 3.67  & $0.38_{-0.25}^{+0.20}$ & $8.18_{-1.13}^{+4.68}$  & 1.14 & 12843
\enddata
\tablecomments{$\gamma$, $P_0$, and $C_{500}$ were varied. The degrees of freedom were 12914.}
\end{deluxetable}

Our best fit model comes from the S13 values of $\alpha$ and $\beta$, and finds $\gamma = 0.38$, $C_{500} = 1.19$, and
$P_0 = 8.18$. The best joint fits, listed in Table \ref{table:m0647_gnfw_fits} to the four sets of $\alpha$ and $\beta$  
differ by $\Delta \Tilde{\chi^2} < 3$. With \citet{young2014} constraining $\gamma = 0.90_{-0.04}^{+0.02}$, it might appear 
that their result is significantly discrepant with our best fit $\gamma=0.61_{-0.11}^{+0.17}$ from the A10 set, even though 
\citet{young2014} used the identical SZE data as we have used in this analysis. 
A crucial distinction in the fitting procedures is the parameter space searched: in
\citet{young2014}, Bolocam is first fit over grid of fixed $\gamma$ values, $k$ is fixed at 1.0, and the 
parameters $C_{500}$ and $P_0$ are
allowed to float. The resultant fits for each $\gamma$ are then fit as-is (nothing allowed to float) to the
MUSTANG map. Thus, the reported error bars reflect a one-parameter search, without the degeneracies between
$C_{500}$, $P_0$, and $\gamma$ folded into it, and do not include the $\chi^2$ values from the Bolocam fit.

Given that MUSTANG is only able to constrain the pressure profile on scales $9\asec < \theta \lesssim 42\asec$, and 
for MACS 0647, $R_{500} = 3.16$\amin and $C_{500}=1.18$ (A10 and S13 value) or $C{500}=1.8$ (N07 and P12 value), 
then $\beta$ should not relate to the slope within 
the scales probed by MUSTANG. It is possible for $\alpha$ to relate to the slope within the scales in question. However,
As $C_{500}$ decrease, especially below 1.0, as is the case in both this work and \citet{young2014}, then $\alpha$ will 
relate less to the slope within the scales probed by MUSTANG. Crucially, that cluster pressure profile steepen with
increasing radius gives rise to the degeneracy observed between $C_{500}$ and $\gamma$ as seen in Figures
\ref{fig:a1835_ci_params} and \ref{fig:m0647_ci_params}.

\begin{figure}[!h]
\begin{center} 
  \includegraphics[width=0.5\textwidth]{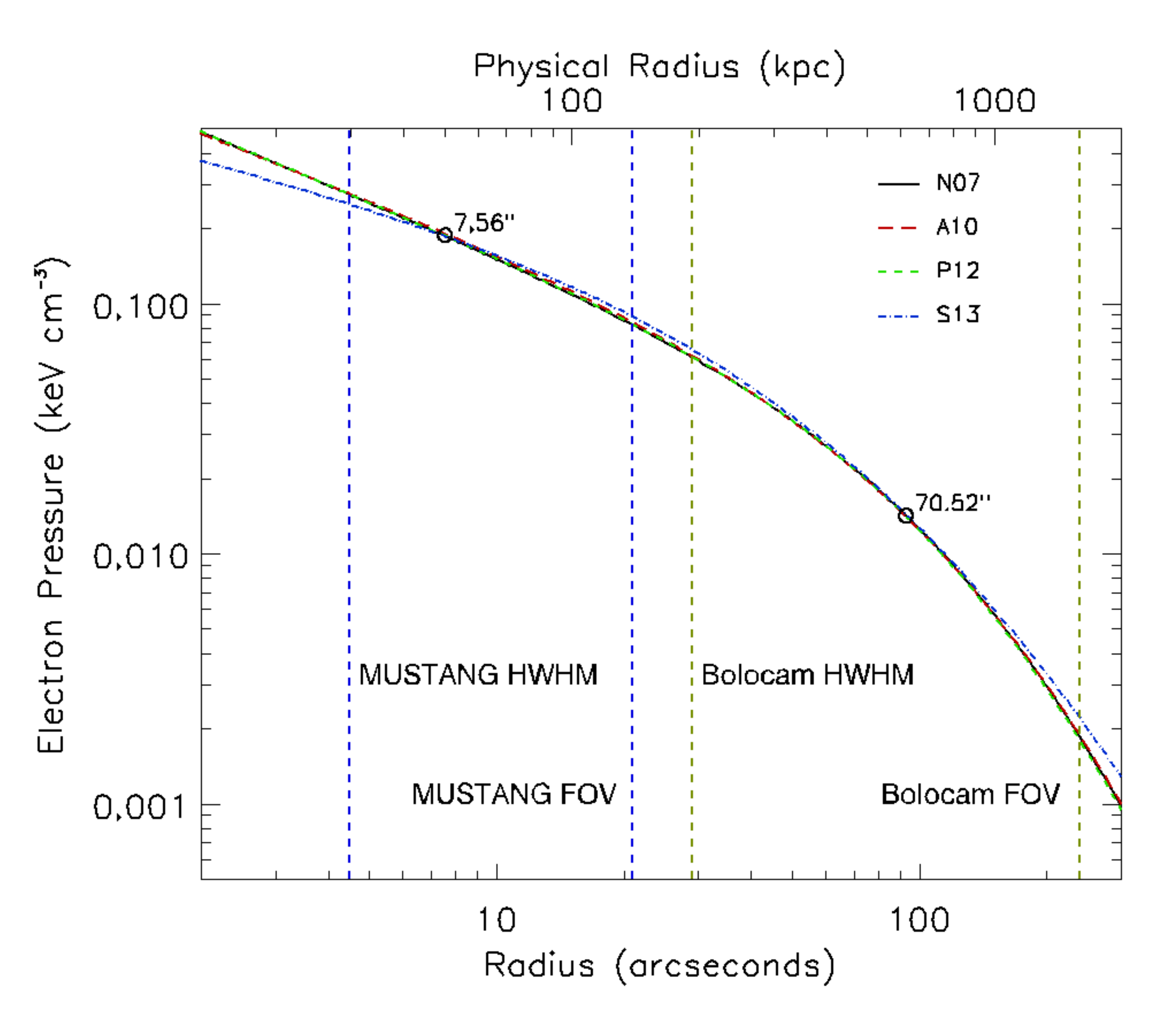}
\end{center} 
  \caption{Best fit pressure profiles for MACS 0647 for the different sets of fixed gNFW $\alpha$ and $\beta$, denoted by the model
   the values are taken from. The circles denote radial ranges where the pressure profiles show (local) minimum scatter.}
  \label{fig:m0647_dpsets}
\end{figure}

\begin{figure}[!ht]
\begin{center} 
  \includegraphics[width=0.5\textwidth]{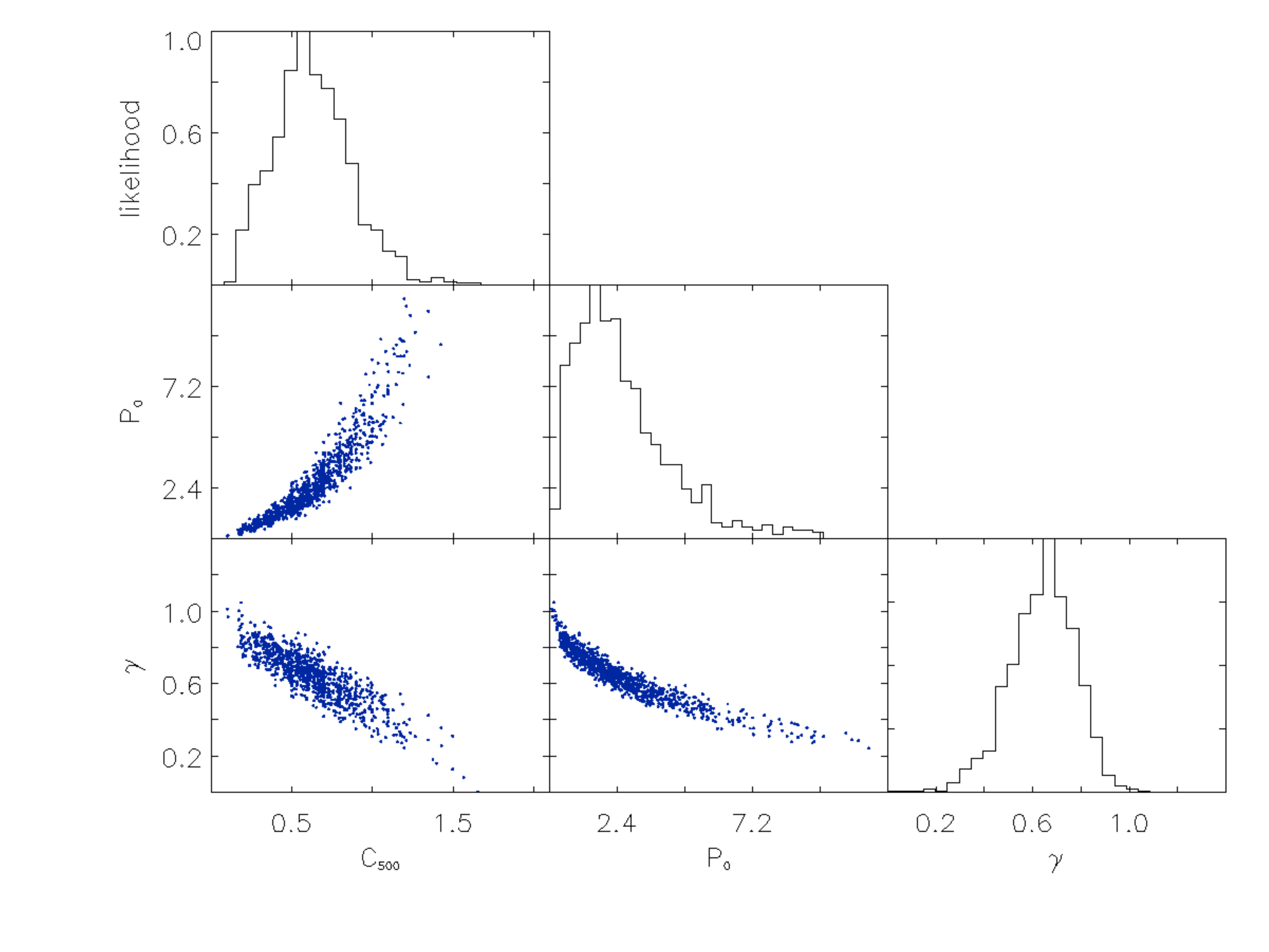}
\end{center} 
  \caption{Results from Monte Carlo fits to 1000 realizations for MACS 0647 with A10 values for $\alpha$ and $\beta$.}
  \label{fig:m0647_density_plot}
\end{figure}

Figure \ref{fig:m0647_ci_params} shows that Bolocam does not place strong constraints on $\gamma$ and $C_{500}$, especially
relative to the joint fit (Table \ref{table:m0647_gnfw_fits}). Specifically, for the A10 set of $\alpha$ and $\beta$, the 
Bolocam finds $C_{500} = 0.66_{-0.38}^{+0.34}$, $\gamma = 0.39_{-0.39}^{+0.49}$, and $P_0 = 17.7_{-8.27}^{+2.76}$.
As in Figure \ref{fig:a1835_dpsets}, we see in Figure \ref{fig:m0647_dpsets} that the best joint fit pressure profiles
from the different gNFW fits of MACS 0647 are in good agreement, and the radii where Bolocam and MUSTANG have the tightest constraints are similar
to the radii of tightest constraints found in Abell 1835. Figure \ref{fig:m0647_density_plot} shows the two-dimensional
confidence intervals over the parameter space searched for MACS 0647.

\subsection{Comparison with ACCEPT}

ACCEPT \citep[][]{cavagnolo2009} utilizes the
\emph{Chandra} Data Archive (CDA) to derive entropy (and pressure) profiles for 239 galaxy clusters. We summarize
how Cavagnolo et al. calculate pressure profiles here.
ACCEPT pressure profiles are calculated as the product of their derived temperature and electron density profiles.
All the ACCEPT data pulled from CDA for ACCEPT were taken with the ACIS detectors \citep{garmire2003}.
Cavagnolo et al. first derive temperature profiles from spectra. 
The spectra are fitted from annuli containing at least 2500 counts after background subtraction. 
Electron density is then calculated from surface brightness profiles using the 0.7-2.0 keV range. This time,
annuli of $5\asec$ are used (pixels are $0.\asec492$). To account for the slight temperature dependence of X-ray
surface brightness, an appropriate temperature for each surface brightness annulus is interpolated from the
radial temperature profile grid.

\begin{figure}[!h]
\begin{center}
  \includegraphics[width=0.5\textwidth]{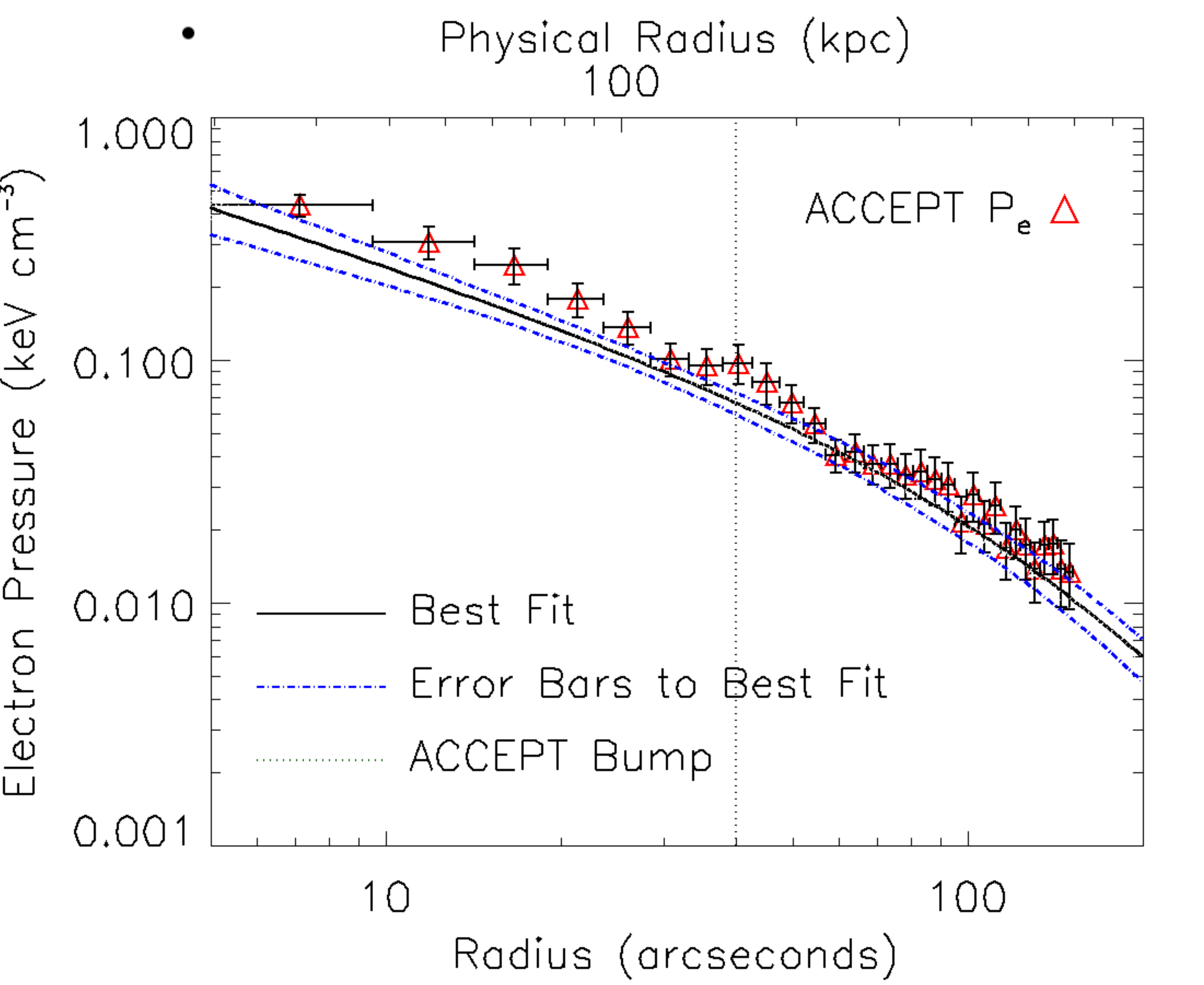}
\end{center} 
  \caption{Pressure profiles for Abell 1835. The dotted lines show the $1 \sigma$ error bars on the SZE
	  determined pressure profile.
    The best joint fit from this work is the solid line. ACCEPT error bars are at the 90\% confidence level.}
  \label{fig:a1835_ppp}
\end{figure}

\begin{figure}[!h]
\begin{center} 
  \includegraphics[width=0.5\textwidth]{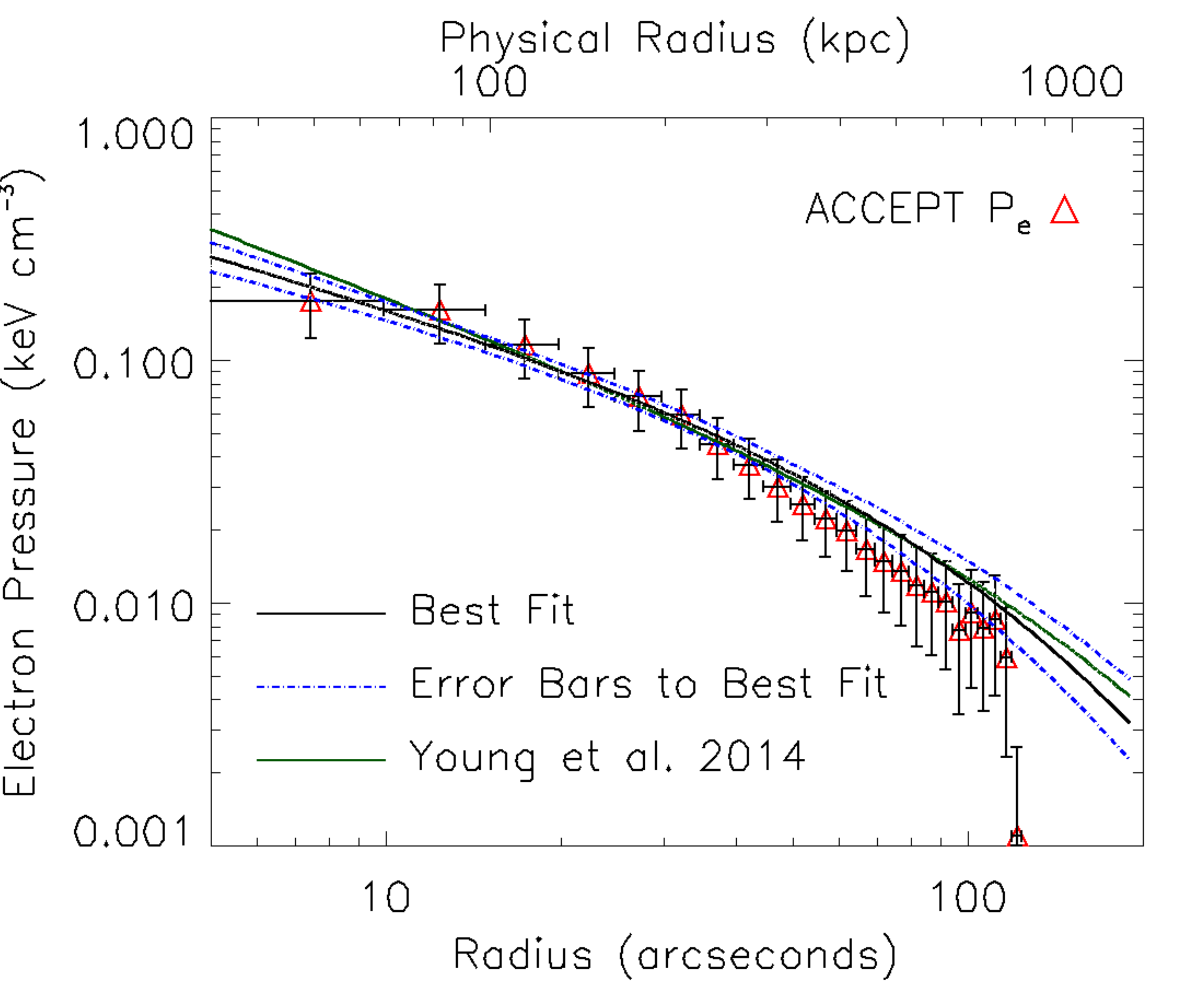}
\end{center} 
  \caption{Pressure profiles for MACS 0647. The best fit (this work) is the solid black line, the dashed lines show the 
    $1 \sigma$ error bars from the joint fit, and the solid dark green is the best fit found in Young et al. 2014.  ACCEPT error bars 
    are at the 90\% confidence level.}
  \label{fig:m0647_ppp}
\end{figure}

We see in Figures \ref{fig:a1835_ppp} and \ref{fig:m0647_ppp} that our joint fits of SZE observations
agree well with ACCEPT data. We note that the ACCEPT profiles show some deviation from a gNFW curve. For Abell
1835, there is a bump in the pressure at $r \sim 40\asec$. In MACS 0647, from $r \sim 15\asec$ to $r \sim 100\asec$,
the ACCEPT pressure profile is almost a pure power law. 

\section{Discussion}
\label{sec:discussion}

Here we have presented an approach to jointly fit models to SZE data from different
instruments. 
With regards to our current sample, we have shown that this approach finds pressure profiles
that are in good agreement with X-ray derived pressure profiles. We choose to vary gNFW
parameters that correspond to the central regions ($P_0$, $C_{500}$, and $\gamma$) given that
we expect our data will be best suited to constraining these, and to avoid encountering the
large degeneracy between all the parameters. 

In Abell 1835 and MACS0647 standard A10 models are detected at 29.7$\sigma$ and 26.3$\sigma$ 
significance respectively.
For Abell 1835, our best fit is $\gamma = 0.36_{-0.36}^{+0.18}$, $C_{500} = 2.29_{-0.68}^{+0.27}$, and 
$P_0 = 19.3_{-8.53}^{+20.4}$ with fixed values of $\alpha = 0.86 $ and $\beta = 3.67$ from S13.
For MACS 0647, our best fit is  $\gamma =0.38_{-0.25}^{+0.20}$, $C_{500} = 1.19_{-0.64}^{+0.54}$, and
$P_0 = 8.18_{-1.13}^{+4.68}$ with fixed values of $\alpha = 0.86$  and $\beta = 3.67$ from S13.
The error bars have been calculated via 1000 Monte Carlo realizations and are marginalized over
the other parameters.
While the spread in fitted parameters appear large (e.g. in Abell 1835, we find $\gamma = 0.36$
with S13 values of $\alpha$ and $\beta$, and we find $\gamma=0.84$ for P12 values of $\alpha$ 
and $\beta$), we note that the fitted pressure profiles are in very good agreement 
(Figures \ref{fig:a1835_dpsets} and \ref{fig:m0647_dpsets}), and the
profiles reveal nodes of least scatter set by MUSTANG and Bolocam. These two nodes
occur roughly where expected: at the geometric mean each instrument's resolution and FOV.

With this approach, we find that contaminants such as point sources can be well modeled and 
their amplitudes are minimally degenerate with cluster pressure models. The pressure profile
constraints do not depend on the MUSTANG beam shape assumed to model point sources. 
Furthermore, we find joint fits with 
MUSTANG and Bolocam allow tight constraints on the pressure profiles over a large range of scales.

Confidence intervals for both MACS 0647 and Abell 1835 show that the constraints are weaker than
indicated from $\Tilde{\chi}^2$. This discrepancy arises because the true covariance matrix for
Bolocam is not diagonal, as assumed for individual fits.

Abell 1835 and MACS 0647 also illustrate the importance of spanning a range of angular scales. While Bolocam
data alone provide good constraints on $\gamma$ and $C_{500}$ for Abell 1835, Bolocam data do not provide precise
constraints on these parameters for MACS 0647. While data quality and quantity may explain some of the difference, 
a critical distinction is that $R_{500}$ occurs at $6.30\arcmin$ for Abell 1835, and $3.16\arcmin$ for MACS 0647, which is
primarily due to Abell 1835 being at lower redshift ($z=0.253$) than MACS 0647 ($z=0.591$). The inner 
pressure profile of Abell 1835 can be constrained by Bolocam, but due to its larger angular extent, MUSTANG
will filter out much ($r > 1\arcmin$) of the profile within $R_{500}$. However, MACS 0647 appears smaller on the sky 
and higher resolution (than $58\asec$) is necessary to constrain the inner pressure profile. Therefore,
the addition of high resolution MUSTANG data becomes crucial for clusters at large redshifts.

A significant improvement to this approach would be to reduce the computational resources required.
Currently, the forward filtering of our MUSTANG models is the most computationally expensive step. As this step
is required for each model created, this restricts how many models we can produce, and thus it restricts
the parameter space we can search. Computing a transfer function for MUSTANG would significantly reduce the total
computational expense; however, the applicability of such a transfer function must be assessed.

We plan to expand this joint fitting technique to a larger sample of clusters, and we note that both Bolocam
and MUSTANG show good internal consistency in their flux calibration. As a result, a single value of $k$ 
should be valid for all of the clusters observed by both instruments, and therefore we expect it can be
tightly constrained using the full sample. This will effectively eliminate one of the free parameters in our fits,
and produce tighter constraints on the gNFW model parameters.

\section*{Acknowledgments}

We thank the anonymous referee whose suggestions have improved this manuscript.
Support for CR was provided through the Reber Fellowship. Support for CR, PK, and AY was provided by 
the Student Observing Support (SOS) program. Support for TM is provided by the National Research Council 
Research Associateship Award at the U.S.\ Naval Research Laboratory. Basic research in radio astronomy
at NRL is supported by 6.1 Base funding. JS was partially supported by a 
Norris Foundation CCAT Postdoctoral Fellowship and by NSF/AST-1313447. Additional funding has been
provided by NSF/AST-1309032.

The National Radio Astronomy Observatory is a facility of the National Science Foundation which is operated
under cooperative agreement with Associated Universities, Inc. The GBT observations used in this paper were
taken under NRAO proposal IDs AGBT09A052, AGBT09C059, AGBT11A009, and AGBT11B001. The assistance of GBT
operators Dave Curry, Greg Monk, Dave Rose, Barry Sharp, and Donna Stricklin. 

The Bolocam observations presented here were obtained form the Caltech Submillimeter Observatory, which,
when the data used in this analysis were taken, was operated by the California Institute of Technology under
cooperative agreement with the National Science Foundation. Bolocam was constructed and commissioned using funds
from NSF/AST-9618798, NSF/AST-0098737, NSF/AST-9980846, NSF/AST-0229008, and NSF/AST-0206158. Bolocam observations
were partially supported by the Gordon and Betty Moore Foundation, the Jet Propulsion Laboratory Research and
Technology Development Program, as well as the National Science Council of Taiwan grant NSC100-2112-M-001-008-MY3.


\bibliographystyle{apj}
\bibliography{mycluster}

\end{document}